\documentclass[aps,prc,groupedaddress,showpacs,twocolumn]{revtex4}
\usepackage{graphicx}
\usepackage{longtable}

\begin{document}


\title{Structure of $^{65,67}$Co studied through the $\beta$ decay of $^{65,67}$Fe and a deep-inelastic reaction}


 \author{D.~Pauwels}
 \email[]{Dieter.Pauwels@fys.kuleuven.be}
 \affiliation{Instituut voor Kern- en Stralingsfysica, K.U. Leuven,
 Celestijnenlaan 200D, B-3001 Leuven, Belgium}
 \author{O.~Ivanov}
 \affiliation{Instituut voor Kern- en Stralingsfysica, K.U. Leuven,
 Celestijnenlaan 200D, B-3001 Leuven, Belgium}
 \author{N.~Bree}
 \affiliation{Instituut voor Kern- en Stralingsfysica, K.U. Leuven,
 Celestijnenlaan 200D, B-3001 Leuven, Belgium}
 \author{J.~B\"{u}scher}
 \affiliation{Instituut voor Kern- en Stralingsfysica, K.U. Leuven,
 Celestijnenlaan 200D, B-3001 Leuven, Belgium}
 \author{T.E.~Cocolios}
 \affiliation{Instituut voor Kern- en Stralingsfysica, K.U. Leuven,
 Celestijnenlaan 200D, B-3001 Leuven, Belgium}
 \author{M.~Huyse}
 \affiliation{Instituut voor Kern- en Stralingsfysica, K.U. Leuven,
 Celestijnenlaan 200D, B-3001 Leuven, Belgium}
 \author{A.~Korgul}
 \affiliation{Institute of Experimental Physics, Warsaw University, ul.Ho\.{z}a 69, 00-681 Warszawa, Poland}
 \author{Yu.~Kudryavtsev}
 \affiliation{Instituut voor Kern- en Stralingsfysica, K.U. Leuven,
 Celestijnenlaan 200D, B-3001 Leuven, Belgium}
 \author{R.~Raabe}
 \affiliation{Instituut voor Kern- en Stralingsfysica, K.U. Leuven,
 Celestijnenlaan 200D, B-3001 Leuven, Belgium}
 \author{M.~Sawicka}
 \affiliation{Instituut voor Kern- en Stralingsfysica, K.U. Leuven,
 Celestijnenlaan 200D, B-3001 Leuven, Belgium}
 \author{I.~Stefanescu}
 \affiliation{Department of Chemistry and Biochemistry, University of Maryland, College Park, Maryland 20742, USA}
\affiliation{Physics Division, Argonne National Laboratory, Argonne, Illinois 60439, USA}
 \author{J.~Van de Walle}
 \affiliation{Instituut voor Kern- en Stralingsfysica, K.U. Leuven,
 Celestijnenlaan 200D, B-3001 Leuven, Belgium}
 \author{P.~Van Duppen}
 \affiliation{Instituut voor Kern- en Stralingsfysica, K.U. Leuven,
 Celestijnenlaan 200D, B-3001 Leuven, Belgium}
 \author{W.B.~Walters}
 \affiliation{Department of Chemistry and Biochemistry, University of Maryland, College Park, Maryland 20742, USA}
\author{R. Broda}
\affiliation{Niewodniczanski Institute for Nuclear Physics, Krakow, PL-31342, Poland}
\author{M.P. Carpenter}
\affiliation{Physics Division, Argonne National Laboratory, Argonne, Illinois 60439, USA}
\author{R.V.F. Janssens}
\affiliation{Physics Division, Argonne National Laboratory, Argonne, Illinois 60439, USA}
\author{B. Fornal}
\affiliation{Niewodniczanski Institute for Nuclear Physics, Krakow, PL-31342, Poland}
\author{A.A. Hecht}
\affiliation{Department of Chemistry and Biochemistry, University of Maryland, College Park, Maryland 20742, USA}
\affiliation{Physics Division, Argonne National Laboratory, Argonne, Illinois 60439, USA}
\author{N. Hoteling}
\affiliation{Department of Chemistry and Biochemistry, University of Maryland, College Park, Maryland 20742, USA}
\affiliation{Physics Division, Argonne National Laboratory, Argonne, Illinois 60439, USA}
\author{A. W\"ohr}
\affiliation{Department of Chemistry and Biochemistry, University of Maryland, College Park, Maryland 20742, USA}
\affiliation{Physics Division, Argonne National Laboratory, Argonne, Illinois 60439, USA}
\author{W. Krolas}
\affiliation{Niewodniczanski Institute for Nuclear Physics, Krakow, PL-31342, Poland}
\author{T. Lauritsen}
\affiliation{Physics Division, Argonne National Laboratory, Argonne, Illinois 60439, USA}
\author{T. Pawlat}
\affiliation{Niewodniczanski Institute for Nuclear Physics, Krakow, PL-31342, Poland}
\author{D. Seweryniak}
\affiliation{Physics Division, Argonne National Laboratory, Argonne, Illinois 60439, USA}
\author{J.R. Stone}
\affiliation{Department of Chemistry and Biochemistry, University of Maryland, College Park, Maryland 20742, USA}
\affiliation{Department of Physics, University of Oxford, OX1 3PU Oxford, United Kingdom}
\author{X. Wang}
\affiliation{Physics Division, Argonne National Laboratory, Argonne, Illinois 60439, USA}
\affiliation{Department of Physics, University of Notre Dame, Notre Dame, Indiana 46556, USA}
\author{J. Wrzesinski}
\affiliation{Niewodniczanski Institute for Nuclear Physics, Krakow, PL-31342, Poland}
\author{S. Zhu}
\affiliation{Physics Division, Argonne National Laboratory, Argonne, Illinois 60439, USA}



\date{\today}

\begin{abstract}
The neutron-rich isotopes $^{65,67}$Fe and $^{65}$Co have been produced at the LISOL facility, Louvain-La-Neuve, in the proton-induced fission of $^{238}$U. Beams of these isotopes have been extracted with high selectivity by means of resonant laser ionization combined with mass separation. Yrast and near-yrast levels of $^{65}$Co have also been populated in the $^{64}$Ni+$^{238}$U reaction at Argonne National Laboratory. The level structure of $^{65}$Co could be investigated by combining all the information from both the $^{65}$Fe and $^{65}$Co $\beta$ decay and the deep-inelastic reaction. The $^{65}$Fe, $^{65}$Co and $^{67}$Fe decay schemes and the $^{65}$Co yrast structure are fully established. The $^{65,67}$Co level structures can be interpreted as resulting from the coexistence of core-coupled states with levels based on a low-energy proton-intruder configuration.
\end{abstract}

\pacs{23.40.-s,23.20.Lv,21.10.-k,27.50.+e}

\maketitle

\section{Introduction}

The region around $Z=28$ and $N=40$ has drawn considerable
interest in nuclear-structure research since the observation in $^{68}$Ni of a $0^{+}$ level as the first excited state \cite{Ber_PL_82} and the high excitation energy of the first excited $2^{+}$ level \cite{Bro_PRL_95}. Both properties were regarded as strong indications for the double-magic character of $^{68}$Ni. However, despite all the additional information that was acquired over the last decade, the specific role of the $N=40$ subshell closure and the $Z=28$ closure on the structure of nuclei around $^{68}$Ni is not yet understood. Especially the occurrence of single-particle and collective phenomena close to $^{68}$Ni deserves further attention.
Therefore, this region remains challenging from both an experimental and a theoretical perspective.

Most early studies were aimed at the structure of the  more easily produced nickel and copper isotopes using in-beam $\gamma$-ray studies \cite{Bro_PRL_95,Paw_NPA_94,Grz_PRL_98}, $\beta$-decay investigations \cite{Fra_PRL_98,Mue_PRL_99,Wei_PRC_99,Mue_PRC_00,Van_PRC_04}, and (d,$^{3}$He) pickup reactions \cite{Zei_PRC_78}. Recently,  more exotic nuclei with $Z \ge 28$ have provided experimental information providing a more direct connection with single-particle and collective configurations, e.g., g-factor measurements of $^{69m}$Cu and $^{67m}$Ni \cite{Geo_JPG_02} and transition probabilities from Coulomb excitation experiments in $^{68}$Ni \cite{Sor_PRL_02,Bre_PRC_08} and $^{67-71,73}$Cu \cite{Ste_PRL_07,Ste_PRL_08}. These results
provide complementary, but often also unanticipated and crucial insights, such as the admixture of $(\mathrm{1p-1h})$ excitations
across $Z=28$ in $^{67m}$Ni \cite{Geo_JPG_02}, the superfluid
character of $^{68}$Ni \cite{Sor_PRL_02} and the collectivity
present in $^{69}$Cu \cite{Ste_PRL_08}. While experiments on the nickel and copper isotopes are currently experiencing a new boost,
information on the cobalt isotopes around and beyond $N=40$ remains scarce. On the other hand, the levels of the lighter odd-mass cobalt isotopes up through $^{63}$Co appear to be readily described by treating the observed structures as resulting from a single $\pi f_{7/2}$ proton-hole coupled to the adjacent even-even nickel cores \cite{Reg_PRC_96}.

Until recently, the known experimental observables for the heavier cobalt nuclei near $^{68}$Ni were isomeric
$\gamma$ decay in $^{66}$Co \cite{Grz_PRL_98}, a $189$-keV
$\gamma$ transition in $^{67}$Co \cite{Sor_NPA_00} and ground state half-life values determined in $^{66-71,73}$Co $\beta$ decay
\cite{Mue_PRC_00,Wei_PRC_99,Mue_PRL_99,Saw_EPJ_04,Raj_SPR_07}.
$^{65}$Co is the heaviest odd-mass cobalt isotope for which, prior
to this work, a level scheme was proposed \cite{Gau_The_05}, albeit one that could not be interpreted. The $^{65}$Co level scheme presented in the present paper results from a study of $^{65}$Fe $\beta$ decay, as well as of deep-inelastic data that were initially dedicated to the yrast structure of $^{64}$Fe \cite{Hot_PRC_06}. Also, the study of the subsequent $^{65}$Co $\beta$ decay turned out to be crucial to establish the $^{65}$Co structure. In previous $\beta$-decay work \cite{Bos_NPA_88}, the $(7/2^{-})$ ground state of $^{65}$Co was reported to directly $\beta$ feed the $1/2^{-}$ state at $1274$~keV, which remained an unsolved issue. In addition to the $^{65}$Fe-$^{65}$Co decay study presented here, the $\beta$ decay of $^{67}$Fe was studied in the same experimental campaign and revealed a $\pi(1p-2h)$ isomer at $492$~keV in $^{67}$Co \cite{Pau_PRC_08}. While in Ref.~\cite{Pau_PRC_08} the isomer was discussed, the present paper is focussing on a detailed description and discussion of the full $^{67}$Fe decay scheme.

There are also theoretical issues. Strong monopole
interactions \cite{Ots_PRL_05} are present between the $\pi f_{7/2}$ orbital and the $\nu f_{5/2}$, $\nu g_{9/2}$ and, according to Refs.~\cite{Cau_EPJ_02,Sor_EPJ_03}, even the $\nu d_{5/2}$ orbitals. However, the interactions are not precisely known and, moreover, they currently cannot all be included simultaneously in the same valence space. Nevertheless, the first excited levels in copper and nickel isotopes are generally described in fair agreement with experiment by large-scale shell-model calculations \cite{Smi_PRC_04} using appropriate effective interactions based on the G matrix \cite{Hjo_PR_95} and modified further with a monopole
correction \cite{Now_The_96}. Experimental spectroscopic-factor values of the first excited $5/2^{-}$ state in the copper
chain determined up to $A=65$, are, however, significantly larger than the calculated values \cite{Smi_PRC_04}. This discrepancy
illustrates the deficiency in either the interactions used or in the restricted valence space. Unfortunately, even more demanding calculations are required when the $\pi f_{7/2}$ orbital is active; i.e., for cobalt nuclei and other $Z < 28$ nuclei. With the hitherto limited valence space and current realistic interactions, large-scale shell model calculations are not yet available for a consistent description of these nuclei.

The isotopes that are studied in this paper, $^{65,67}$Co, provide
important information on the effective interactions in
the cobalt isotopes adjacent to $^{68}$Ni. Consequently, they
also form a good testing ground for the potential magic character of the $^{68}$Ni core. The isotopes form a bridge between the spherical nickel isotopes and the region of deformation below $Z=28$ that is observed to set in gradually in excited states of $^{66}$Fe
\cite{Han_PRL_99,Adr_PRC_08} and is proposed for the
$^{64}$Cr ground state \cite{Sor_EPJ_03,Adr_PRC_08,Aoi_NPA_08}. Because the onset of deformation below $Z=28$ is understood only qualitatively, it is not clear a priori how the cobalt isotopes are behaving, since the deformation mechanism depends critically on the values of the $N=40$ and $N=50$ gaps \cite{Cau_EPJ_02}. It is known that the $N=40$ subshell closure is already weak in $^{68}$Ni ($\Delta E = S_{2n}(Z=28,N=40)-S_{2n}(Z=28,N=42) = 1.71(3)$ MeV) \cite{Rah_EPJ_07} and decreases further in $^{67}$Co ($\Delta E = 1.0(6)$ MeV) \cite{Aud_NPA_03}, although the error bar is large in the latter case. Under the influence of the tensor interaction between protons in the $\pi f_{7/2}$ orbital and neutrons in the $\nu f_{5/2}$ and $\nu g_{9/2}$ orbitals \cite{Ots_PRL_06}, the closure is expected to decrease even further for the $N=40$ isotones with lower Z. The $N=50$ energy shell gap in $^{78}$Ni is not experimentally determined yet, but the systematics of the $N=50$ isotones with $Z=31-40$ indicate the persistence of this gap towards nickel \cite{Hak_PRL_08}. Nuclear structure of the cobalt isotopes was until recently known up to $N=37$ and the structure of all of the nuclei can be interpreted as originating from a $\pi f_{7/2}^{-1}$ proton hole coupled to its adjacent nickel neighbor. Moreover, the dominant low-energy structure of $^{67,69}$Ni and $^{68-70}$Cu can be explained as due to a coupling with excited levels of the $^{68}$Ni core \cite{Oro_NPA_00}. It came, therefore, as a surprise that the first excited level in $^{67}$Co arises from excitations across $Z=28$ \cite{Pau_PRC_08}. The complementarity of the $\beta$-decay and deep-inelastic experiments allows to interpret the $^{65}$Co structure. Also, a more detailed $^{67}$Fe $\beta$-decay scheme is presented and will be discussed extensively.

\section{Experimental setup}

\subsection{$\beta$ decay at LISOL}

\begin{table*}[htb]
\begin{center}
\caption{List of the different data sets with indication of the
mass $A$, the laser resonance, the cycle (beam ON/beam OFF/number
of cycles per tape move), the total data-acquiring time $\Delta t$, the production rate $P$ of the listed
isotope and the parameters of equation \ref{eq:gamma_eff}, which determine the $\gamma$-ray photo-peak efficiency.\label{tbl:DataSets}}
\begin{footnotesize}
\begin{ruledtabular}
\begin{tabular}{ccccccccccc}
 Data set & A  & Lasers & Cycle & $\Delta t$ ($h$) & Isotope & $P$ (at$/\mu C$)  & $C_{1}$ & $D_{1}$ & $C_{2}$ & $D_{2}$ \\
\hline
 I   & 65 & Fe  & $2.4$s/$2.4$s/$5$ & 49.9  & $^{65}$Fe & 1.23(8)    & $85(11) \cdot 10^{-5}$ & $-792(19) \cdot 10^{-3}$ & $800(1000)$ & $2.16(29)$ \\
     &    &     &                   &     & $^{65}$Fe$^{m}$ & 1.20(17)    & & & & \\
\hline
 II  & 65 & Co  & $2.4$s/$2.4$s/$5$ & 28.6  & $^{65}$Co & 6.6(16)
 & $85(11) \cdot 10^{-5}$ & $-792(19) \cdot 10^{-3}$ & $800(1000)$ & $2.16(29)$ \\
\hline
 III & 65 & off & $2.4$s/$2.4$s/$5$ & 4.4   & $^{65}$Co & $< 1.1$
 & $85(11) \cdot 10^{-5}$ & $-792(19) \cdot 10^{-3}$ & $800(1000)$ & $2.16(29)$ \\
 &        &     &                   &      & $^{65}$Fe & $< 0.05$
 & & & & \\
 &        &     &                   &     & $^{65}$Fe$^{m}$ & $< 0.05$
 & & & & \\
\hline
 IV  & 67 & Fe  & $1.4$s/$1.6$s/$3$ & 9.9  & $^{67}$Fe & 1.37(13)
 & $3.1(4) \cdot 10^{-3}$ & $-58(25) \cdot 10^{-2}$  & 0  &   \\
\hline
 V   & 67 & off & $1.4$s/$1.6$s/$3$ & 5.3  & $^{67}\mathrm{Co} + ^{67}$Co$^{m}$ & 0.40(8)
 & $3.1(4) \cdot 10^{-3}$ & $-58(25) \cdot 10^{-2}$  & 0  &   \\
 &        &     &                   &      & $^{67}$Fe & $< 0.02$
 & & & & \\
\hline
 VI  & 67 & Fe  & $10$s/$0$s/$1$    & 53.9 & $^{67}$Fe & 0.45(13)
 & $1.7(2) \cdot 10^{-3}$ & $-65(2) \cdot 10^{-2}$  & 0  &   \\
\hline
 VII & 67 & off & $10$s/$0$s/$1$    & 8.2  & $^{67}$Co$^{m}$ & 0.51(14)   & $1.7(2) \cdot 10^{-3}$ & $-65(2) \cdot 10^{-2}$  & 0  &   \\
 &        &     &                   &      & $^{67}$Co & 0.25(7)
 & & & & \\
 &        &     &                   &      & $^{67}$Fe & $< 0.02$
 & & & & \\
\end{tabular}
\end{ruledtabular}
\end{footnotesize}
\end{center}
\end{table*}

Short-lived $^{65,67}$Fe and $^{65,67}$Co isotopes have been
produced at the LISOL facility \cite{Kud_NIM_03,Fac_NIM_04} installed at the
Cyclotron Research Center (CRC) at Louvain-La-Neuve (Belgium) with the $30$-MeV
proton-induced fission reaction on $^{238}$U. The two $10 \
\mathrm{mg}/\mathrm{cm}^2$ thin $^{238}$U targets were placed
inside a gas catcher in order to stop and thermalize the recoiling fission products in an argon buffer gas with $500$ mbar pressure.
As the fission products, dragged by the argon flow, come close to
the exit hole of the gas cell, they are irradiated by two
excimer-pumped dye lasers that resonantly ionize the desired
element. The ions leaving the cell are transported through a
SextuPole Ion Guide (SPIG) \cite{Van_NIM_97} to a high-vacuum
environment, where they are accelerated over a potential difference of $40$~kV. After separation according to their mass-to-charge ratio $A/Q$, the ions are implanted into a detection tape surrounded by three thin plastic $\Delta E$ $\beta$ detectors and
two MINIBALL $\gamma$-detector clusters \cite{Ebe_NPA_01}.

Important features of the detection setup, described in
\cite{Pau_NIM_08}, are the MINIBALL's granularity and the
data acquisition through digital electronics.
The granularity reduces substantially the loss in photo-peak efficiency due to true $\gamma$ summing. In the data acquisition, all $\beta$ and $\gamma$ events get an absolute time stamp by a
$40$-MHz clock, so that no timing information gets lost. The
$\beta$-gated $\gamma$ spectra contain the $\gamma$ events that
occurred in a prompt time window of $350$~ns after a $\beta$ event.
To suppress the true summing of $\beta$ particles with
$\gamma$ rays in the germanium crystals prompt $\gamma$ events are
vetoed if the $\beta$ event occurred at the same side of the
detection setup \cite{Wei_NIM_99}. The combination of digital
electronics and high selectivity by the laser ion source coupled to
mass separation offers the possibility to correlate single
$\gamma$ events and/or $\beta$-gated $\gamma$ events with each
other up into the seconds time range \cite{Pau_NIM_08}, which turned out to be crucial to disentangle the $A=67$ decay scheme from iron down to nickel \cite{Pau_PRC_08}.

For half-life determinations, data were acquired in a cycle
where, in a first period, the cyclotron beam was on and the mass
separator open, followed by a period where the beam was switched off
and the separator closed. After a fixed number of such cycles, the
implantation tape was moved in order to remove long-lived daughter
and contaminant activities. The cycles used are specified in
Table~\ref{tbl:DataSets}, which also contains production rates
at the exit hole of the gas cell as obtained from the observed $\gamma$ intensities. Despite the high purity of the argon buffer gas (at the ppb-level) and the careful preparation of the gas cell,
the production yield is very sensitive to the impurity
level of the buffer gas. Due to slightly different conditions, the $^{67}$Fe production rate was a factor of $3$ higher in data set IV than was the case in data set VI.

The $\gamma$-energy and efficiency calibrations were performed by placing standard $^{133}$Ba, $^{137}$Cs, $^{152}$Eu and $^{60}$Co sources at the implantation spot of the detection tape as well as by on-line implantation of $^{90}$Rb (with intense $\gamma$ lines up to
$3.317$~MeV) and of $^{142}$Ba isotopes, which were produced in
large amounts. The activity could be calculated from the
$\beta$ rate, but at mass $A=142$ the large production of $^{142}$Cs
isotopes had to be coped with. Advantage was taken of the large
difference in half lives between $^{142}$Cs ($T_{1/2}=1.7$~s) and
$^{142}$Ba ($T_{1/2}=10.7$~min) by acquiring data in a
$600\mathrm{s}/600\mathrm{s}/1$ implantation-decay cycle and by not
considering the implantation period and the first $10$~s of the
decay period. The latter notation denotes a cycle of $600$~s implantation and $600$~s decay and the implantation tape is moved after $1$ cycle. The $\gamma$ photo-peak efficiencies $\varepsilon_{\gamma}$ (in \%) were fitted by the function
\begin{equation}
\varepsilon_{\gamma} = \frac{1}{C_{1} E_{\gamma}^{-D_{1}} + C_{2} E_{\gamma}^{-D_{2}}},
\label{eq:gamma_eff}
\end{equation}
where $E_{\gamma}$ denotes the $\gamma$ energy in units of keV and the parameters $C_{1}$, $C_{2}$, $D_{1}$ and $D_{2}$ are given in Table~\ref{tbl:DataSets} for the respective data sets. Note that for data sets IV-VII the second term in the denominator, which fits the low-energy behavior, was fixed to $0$, since there are no $\gamma$ transitions with low energy in the $^{67}$Fe decay.

\subsection{Deep-inelastic reaction at ANL}

Complementary information for the $^{65}$Co level structure was extracted from an experiment performed at Argonne National Laboratory. The prime objective of that experiment was to study excited levels in the neutron-rich iron isotopes populated in deep-inelastic reactions of a $430$-MeV $^{64}$Ni beam with a $55$~mg/cm$^2$ isotopically-enriched $^{238}$U target \cite{Hot_PRC_06,Hot_PRC_08}. The uranium target was located in the center of  the Gammasphere array \cite{Lee_NPA_90} consisting of $100$ Compton suppressed HPGe detectors. The nickel beam was produced in bunches separated by 82 ns, but for the deep-inelastic experiment reported here, only one out of five pulses was allowed to hit the Gammasphere target. This resulted in prompt bursts separated by a 410-ns gap within which delayed $\gamma$-ray decays emitted by the reaction products could be studied.

Events were recorded on the basis of three-fold or higher-order coincidences. Data were sorted offline into single $\gamma$ spectra, $\gamma$-$\gamma$ matrices and $\gamma$-$\gamma$-$\gamma$ cubes. More details about the sorting procedure are given in Refs. \cite{Hot_PRC_06,Hot_PRC_08}. Four types of coincidence cubes PPP (prompt-prompt-prompt), DDD (delayed-delayed-delayed), PDD (prompt-delayed-delayed) and PPD (prompt-prompt-delayed) were created by selecting different $\gamma$-ray times with respect to the prompt beam bursts. The PPP cube was obtained by selecting only the events recorded within $\pm$20 ns of the beam burst. These events correspond to $\gamma$ rays emitted by excited levels populated directly in the deep-inelastic reaction. The DDD cube was constructed by selecting only the delayed events (but with the three $\gamma$ rays within a prompt coincidence window of $40$ ns) acquired during the beam-off period. These events consist of $\gamma$ rays emitted by isomeric states ($10$~ns to $10 \ \mu \mathrm{s}$ range) and/or $\beta$-delayed transitions. The PDD and PPD cubes were built by combining the prompt and delayed events and were used to identify transitions above or below isomeric levels. An example of the latter is described in Ref.~\cite{Hot_PRC_08}.


\section{Results}

\subsection{$^{65}$Co decay to $^{65}$Ni}\label{par:65Ni}

\begin{figure*}
\centering
\includegraphics[width=\linewidth]{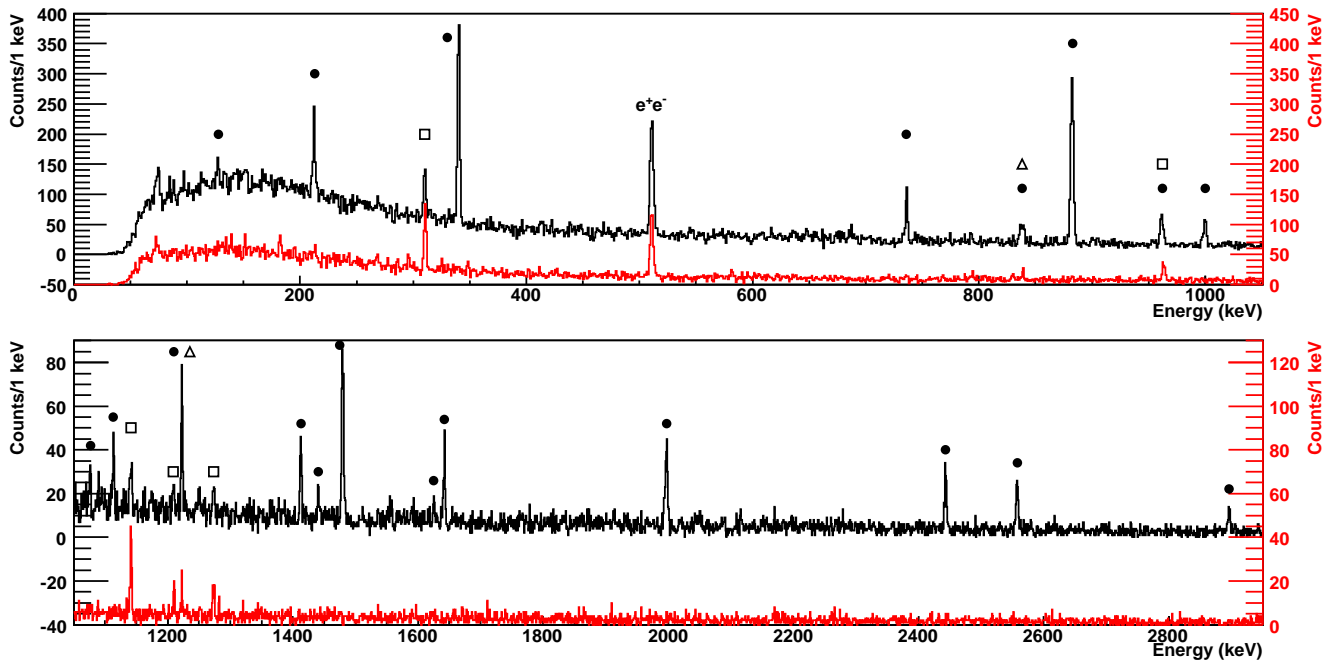}
\caption{(Color on-line) The vetoed $\beta$-gated $\gamma$ spectrum with the lasers tuned to ionize iron and cobalt in black and red, respectively. The lines from $^{65}$Fe to $^{65}$Co decay are indicated by full circles, those from $^{65}$Co to $^{65}$Ni decay by open squares and contaminant lines by open triangles. Three doublets are present: at $837$
(of $^{65}$Fe and $^{130m}$Sb decay), $962$ (of $^{65}$Fe and
$^{65}$Co decay) and $1223$~keV (of $^{65}$Fe and $^{98}$Y decay).
The $^{130m}$Sb ions and $^{98}\mathrm{Y}\mathrm{O}_{2}$ molecules
were able to reach the detection tape in a doubly-charged form (see text for details).}
 \label{fig:65Fe_spectrum}
\end{figure*}

Fig.~\ref{fig:65Fe_spectrum} presents the vetoed, $\beta$-gated $\gamma$ spectra of data set I with the lasers tuned on the iron
resonance in black and of data set II with the lasers on the cobalt resonance in red. Using the same cycle, data with the lasers
switched off have also been collected (data set III), only showing a
line at $511$~keV in the vetoed, $\beta$-gated $\gamma$ spectrum. In data set II, significant contaminant lines are observed at $838$ and $1223$~keV, originating from the $\beta$ decay of $^{130m}$Sb and $^{98}$Y, respectively. $^{130m}$Sb and the molecule $^{98}$YO$_2$ are able to reach the detection tape in a doubly-charged state. In
Fig.~\ref{fig:65Fe_spectrum}, the full circles indicate lines from the $^{65}$Fe to $^{65}$Co decay, open squares from $^{65}$Co to $^{65}$Ni and open triangles from contaminant $\beta$ decay. The observed transitions, count rates and relative intensities in $^{65}$Ni from data set II are listed in Table \ref{tbl:65ColinesNi}, while this information for the transitions observed in $^{65}$Ni and $^{65}$Co from data set I is summarized in Tables \ref{tbl:65FelinesNi} and
\ref{tbl:65FelinesCo}, respectively.

\begin{table}[hb]
\begin{center}
\caption{Transitions in $^{65}$Ni from $^{65}$Co $\beta$ decay
(data set II) are indicated by their energy $E$~(keV), the
corresponding off-resonant subtracted $\beta$-gated peak count rate $A_{\gamma}$ and transition intensity $I_{rel}$ relative to the $1141.1$-keV transition ($100 \ \%$). Multiply $I_{rel}$ by 0.027 (7) to get absolute intensities. The $\gamma$-ray energies of coincident events are listed in the last column with the number of observed $\beta$-$\gamma$-$\gamma$ coincidences between brackets.\label{tbl:65ColinesNi}}
\begin{footnotesize}
\begin{ruledtabular}
\begin{tabular}{cccc}
 $E$ (keV)    & $A_{\gamma}$ (cts/h)
                                  & $I_{rel}$ (\%)
                                            & Coincident $\gamma$-events \\
\hline
 63.4 (4)   & 0.9 (2) & 65 (18)\footnote{The relative intensity also includes the correction for electron conversion.} & 1210(3) \\
 310.4 (1)  & 9.2 (7) & 82 (11) & 963(16) \\
 963.4 (2)  & 3.7 (5) & 79 (13) & 310(16) \\
 1141.1 (2) & 4.2 (4) & 100     & -  \\
 1210.6 (2) & 1.5 (3) & 39 (9)  & 63(3) \\
 1273.2 (3) & 1.6 (3) & 42 (9)  & - \\
\end{tabular}
\end{ruledtabular}
\end{footnotesize}
\end{center}
\end{table}

\begin{figure}
\centering
\includegraphics[width=\linewidth]{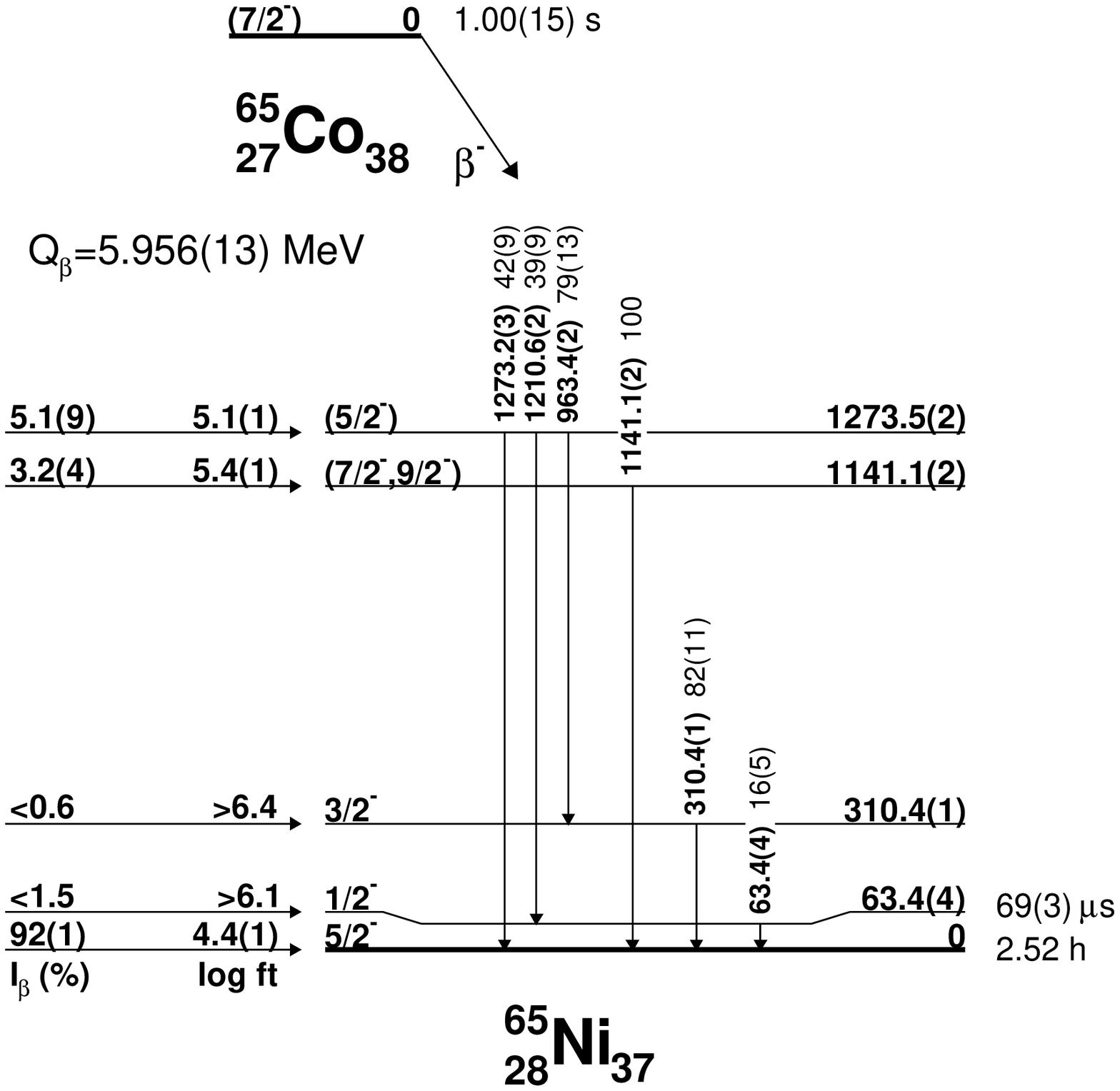}
 \caption{\small{The $^{65}$Co decay scheme into $^{65}$Ni. Spin and parities without brackets are taken from \cite{Fly_PRL_78}, while those within brackets are deduced from the present $\beta$-decay study.}}
 \label{fig:65Ni_Levels}
\end{figure}

A decay scheme for $^{65}$Co was proposed already in earlier
$\beta$-decay work \cite{Bos_NPA_88}. Prompt
$\beta$-$\gamma$-$\gamma$ coincidences (see Table \ref{tbl:65ColinesNi}) confirmed that the $310$- and
$963$-keV transitions are in cascade. Prompt $\beta$-$\gamma$ events
coincident with delayed $63$-keV $\gamma$ rays in a $1$-$150 \ \mu$s time window after the $\beta$ event confirmed that
the $1210$-keV transition feeds an isomeric state with a $63$-keV
excitation energy. Out of the total $63$-keV decays, $22 \ \%$ of
the activity is missed due to the limited time window of $1$ to
$150$~$\mu$s for a $69$~$\mu$s half-life \cite{NNDC} and $75 \ \%$ due to internal electron conversion. The half-life effect is taken into account for its $\gamma$ intensity, as shown in the deduced $^{65}$Co decay scheme of Fig.~\ref{fig:65Ni_Levels}. The relative intensity of the $63$-keV transition, as shown in Table \ref{tbl:65ColinesNi}, also includes the correction for electron conversion, which is necessary to extract the $\beta$ branching towards the $63$-keV level in $^{65}$Ni. No coincidence relationships were observed for the lines of $1273$ and $1141$~keV, confirming their placement as ground-state transitions \cite{Bos_NPA_88}. From comparing the off-resonant subtracted $\beta$ activity with the total $\gamma$ activity a strong ground state feeding of $91.7(8) \ \%$ was deduced, which is consistent with Ref.~\cite{Bos_NPA_88}.

\begin{table}[hb]
\begin{center}
\caption{Transitions in $^{65}$Ni following $^{65}$Fe mother-daughter $\beta$ decay (data set I) are indicated by their energy $E$~(keV), the corresponding off-resonant subtracted peak count rate $A_{\gamma}$, transition intensity $I_{rel}$ relative to the $882.5$-keV transition ($100 \ \%$) and the ratio of relative transition intensities $\frac{I_{rel}(II)}{I_{rel}(I)}$ from data set II and I.\label{tbl:65FelinesNi}}
\begin{footnotesize}
\begin{ruledtabular}
\begin{tabular}{cccc}
 $E$ (keV)    & $A_{\gamma}$ (cts/h)
                       & $I_{rel}$ (\%)
                              & $\frac{I_{rel}(II)}{I_{rel}(I)}$\\
\hline
 63.4 (4)   & 0.50 (12) & 10 (3)    &  6 (2) \\
 310.4 (1)  & 4.1 (5)   & 10.9 (14) &  8 (2) \\
 963.4 (2)  & 1.9 (5)   & 12 (3)    &  7 (2) \\
 1141.1 (2) & 1.4 (4)   & 12 (2)    &  8 (2) \\
 1210.6 (2) & 1.0 (5)   & 8 (2)     &  5 (2) \\
 1273.2 (3) & 1.0 (4)   & 8 (2)     &  6 (2) \\
\end{tabular}
\end{ruledtabular}
\end{footnotesize}
\end{center}
\end{table}

All observed $\gamma$ transitions following $^{65}$Co $\beta$ decay exhibit the same growing-in and decay behavior. A half-life of $1.00(15)$~s was deduced for the $^{65}$Co ground state from a single-exponential fit of the time-dependent summed intensity of these $\beta$-gated $\gamma$ lines during the beam-off period. This value is slightly smaller, but certainly not significantly different from the previously determined half-life values of $1.14 (3)$ s \cite{Bos_NPA_88} and $1.25 (5)$ s \cite{Run_NPA_85}. The decay behavior of $\beta$ and $\gamma$ rays was fitted in the former work, whereas only single $\beta$ rays were fitted in the latter.

The deduced $^{65}$Co decay scheme of Fig.~\ref{fig:65Ni_Levels} does not contain the previously assigned $340$-, $384$- and $882$-keV  transitions \cite{Bos_NPA_88}. The $340$- and $882$-keV $\gamma$ rays follow the $\beta$ decay of $^{65}$Fe, see section \ref{sec:65Co},
while the expected $384$-keV peak integral in the $\beta$-$\gamma$ spectrum would be $17(2)$ as deduced from the $\gamma$ intensities given in Ref.~\cite{Bos_NPA_88} and the $310$-keV peak integral. No $384$-keV peak is observed in the spectrum, but this is still consistent within $2\sigma$ with Ref.~\cite{Bos_NPA_88}, due to the background conditions.

The structure of $^{65}$Ni has also been studied in $(t_{pol},d)$ transfer reaction experiments \cite{Fly_PRL_78} from which spin and parities have been deduced for various levels. The $5/2^{-}$ ground state is strongly fed by direct Gamow-Teller $\beta$ decay, which is consistent with an expected $(7/2^{-})$ ground state of $^{65}$Co. The $1141$-keV level, for which spin and parity quantum numbers were not yet assigned, is strongly fed. Based on a $(7/2^{-})$ $^{65}$Co ground state and a lack of $\gamma$ deexcitation to the low-spin states, a spin and parity of $(7/2^{-})$ or $(9/2^{-})$ can be expected. Also, the state at $1274$~keV gets significant $\beta$ feeding (log $ft=5.1(1)$), and a subsequent strong $\gamma$ decay is observed to the $5/2^{-}$ ground state, the $1/2^{-}$ state at $63$~keV and the $3/2^{-}$ state at $310$~keV. The observed $\gamma$ decay to the $63$-keV, $1/2^{-}$ state rules out $(7/2^{-})$ and $(9/2^{-})$ assignments and leaves $(5/2^{-})$ as the only possibility. This is, however, in disagreement with the previously assigned spin and parity of $1/2^{-}$ in Ref.~\cite{Fly_PRL_78}.

Missing $\gamma$-ray activity from higher lying states feeding the $1274$-keV level seems unlikely due to the relatively small $\beta$-endpoint energy of $5.956(13)$~MeV. Furthermore, the $1/2^{-}$ assignment is not consistent with $^{64}$Ni$(n,\gamma)$ studies. This reaction populates a $1/2^{+}$ level in $^{65}$Ni at $6098$~keV \cite{Coc_PRC_72}, which decays predominantly by high-energy $E1$ transitions to $1/2^{-}$ and $3/2^{-}$ levels. The $(n,\gamma)$ work clearly establishes four $1/2^{-}$ and $3/2^{-}$ states at low energy (at $64$, $310$, $692$ and $1418$~keV). None of these levels are fed directly in the $^{65}$Co $\beta$ decay. Moreover, the $6098$-keV level does not populate directly the $5/2^{-}$ ground state nor the $1141$- and $1274$-keV states, which indicates that the latter two states have spin $J=5/2$ or greater. In the event that a $1/2^{-}$ assignment to the $1274$-keV level would have been correct, the strong $\beta$ feeding would have to originate from a low-spin, $\beta$-decaying isomer in $^{65}$Co. Penning-trap mass measurements, however, could not identify a $\beta$-decaying isomer in $^{65}$Co \cite{Blo_PRL_08}.

\subsection{Level structure in $^{65}$Co}\label{sec:65Co}

\subsubsection{$\beta$-decay study of $^{65}$Fe at LISOL}

Peaks enhanced in the data with the lasers tuned to the iron resonance and absent in the data with the lasers on the cobalt resonance, can be unambiguously assigned to the $^{65}$Fe $\beta$ decay feeding excited states in $^{65}$Co. Hence, the data provide clear evidence that the intense lines at $340$ and
$883$~keV, previously assigned to the $^{65}$Co decay
\cite{Bos_NPA_88}, occur in fact in the $^{65}$Fe $\beta$ decay. At
$838$ and $1223$~keV, the contaminant activity from the doubly-charged $^{130m}$Sb ions and $^{98}$YO$_2$ molecular ions form a doublet with the iron lines. The relative $\gamma$ intensities of the iron lines were obtained by subtraction of the off-resonant contribution, which could be deduced from data set II. The iron line at $961$~keV also forms a doublet with the $963$-keV transition from $^{65}$Co decay. Hence, the contribution of the cobalt line is obtained from the $310$-keV peak integral and the relative $\gamma$ intensities of the $310$- and $963$-keV $\gamma$ rays, as deduced in the $^{65}$Co decay, see Table \ref{tbl:65ColinesNi}. It was checked that the $\gamma$ intensities of both the $1211$-keV and the $1273$-keV transitions relative to the $310$-keV line in data sets I and II are in agreement.

\begin{figure}
\centering
\includegraphics[width=\linewidth]{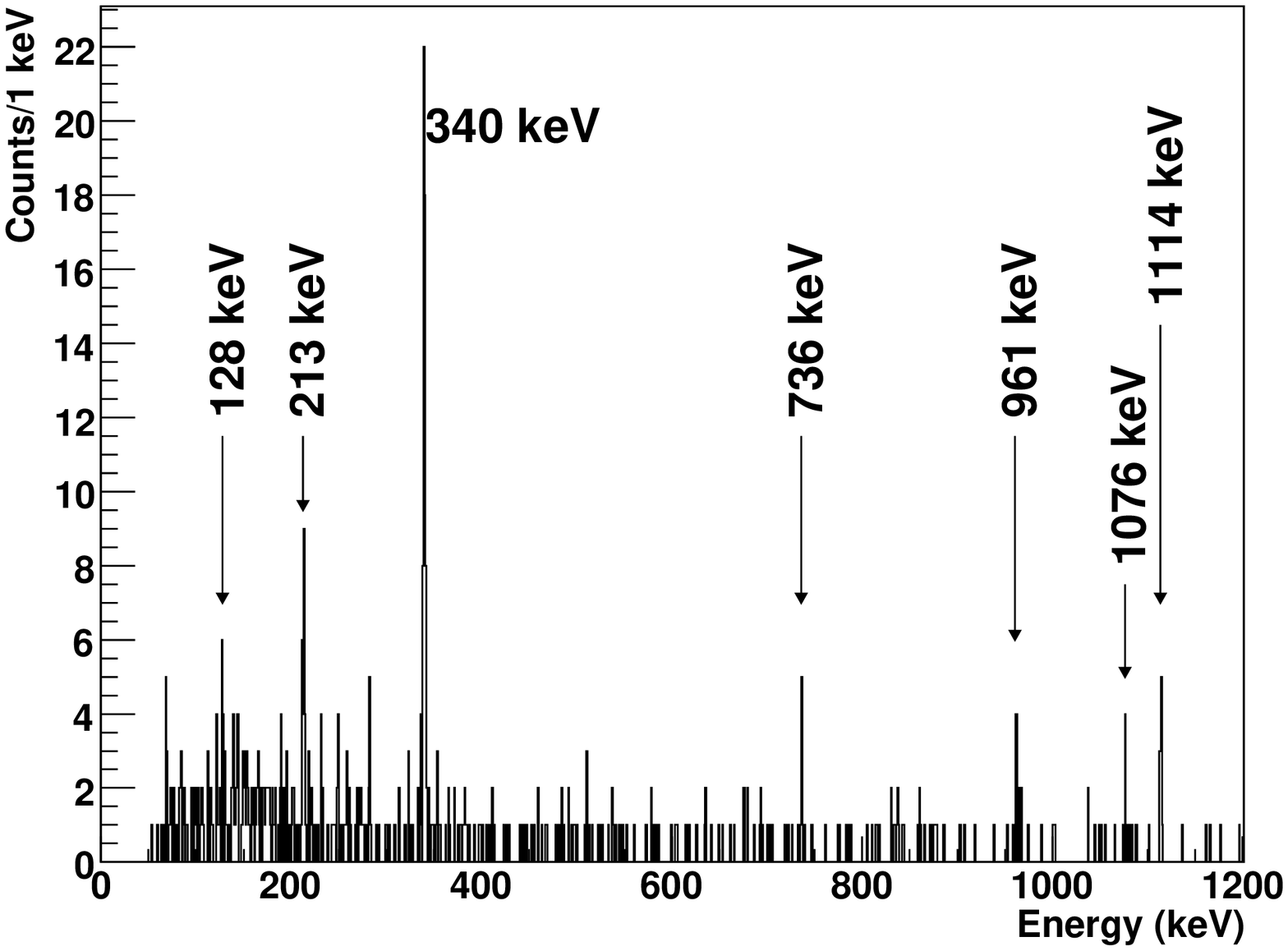}
 \caption{\small{The $\gamma$-ray spectrum of prompt coincidence events with $\beta$-gated, $883$-keV events (data set I).}}
 \label{65Fe_b882g}
\end{figure}

\begin{table*}[htb]
\begin{center}
\caption{Transitions in $^{65}$Co from $^{65}$Fe $\beta$ decay
(data set I) are indicated by their energy $E$~(keV), the
corresponding off resonant subtracted peak count rate $A_{\gamma}$ and transition intensity $I_{rel}$ relative to the $882.5$-keV transition ($100 \ \%$). Multiply $I_{rel}$ by 0.20 (6) to get absolute intensities. The $\gamma$-ray energies of coincident events are listed in the last column with the number of observed $\beta$-$\gamma$-$\gamma$ coincidences between brackets.\label{tbl:65FelinesCo}}
\begin{footnotesize}
\begin{ruledtabular}
\begin{tabular}{cccc}
 $E$ (keV)    & $A_{\gamma}$ (cts/h)
                        & $I_{rel}$ (\%)
                                    & Coincident $\gamma$-events \\
\hline
 127.6 (3)  & 1.7 (6)   & 3.4 (12)  & 213(8), 736(5), 883(5), 961(2) \\
 212.5 (2)  & 5.2 (6)   & 11.1 (13) & 128(7), 736(3), 883(17), 1089(7)\\
 340.07 (6) & 16.5 (7)  & 47 (2)    & 736(27), 774(6), 883(53), 961(8)\\
 413.0 (10) & -         & 2.6 (12)  & 1000(4)\\
 736.1 (10) & 4.3 (4)   &   22 (2)  & 340(28), 883(8), 1223(7)\\
 774.0 (10) & -         &  6 (4)    & 340(5), 883(2), 1223(1)\\
 836.6 (2)  & 2.6 (5)   &   14 (3)  & 1642(4)\\
 864.0 (10) & -         &  1.8 (10) & 213(6), 883(1)\\
 882.50 (9) & 17.0 (6)  &   100     & 128(7), 213(20), 340(53), 736(6), 961(9), 1076(5), 1114(12)\\
 960.5 (2)  & 1.4 (5)   &  9 (3)    & 310(6), 340(10), 883(8) \\
 999.7 (3)  & 2.7 (4)   &   18 (2)  & 413(4), 1480(4) \\
 1076.2 (3) & 1.2 (3)   &  8.3 (18) & 883(6) \\
 1088.7 (6) & 0.6 (2)   &  3.9 (13) & 213(6), 883(2)\\
 1113.5 (3) & 2.1 (3)   &  15 (2)   & 883(12)\\
 1222.7 (2) & 3.1 (4)   &  23 (3)   & 736(7), 774(1), 961(1)\\
 1412.5 (2) & 2.6 (3)   &  22 (3)   & 1480(7)\\
 1441.1 (4) & 0.9 (3)   &  8 (2)    & - \\
 1479.5 (2) & 5.3 (4)   &  47 (3)   & 1000(5), 1413(7) \\
 1625.5 (4) & 0.7 (2)   &  7 (2)    & - \\
 1641.9 (3) & 2.4 (3)   &  23 (3)   & 837(5) \\
 1996.6 (4) & 3.9 (3)   &  44 (4)  &  - \\
 2443.3 (4) & 2.4 (2)   &  31 (3)  &  - \\
 2557.5 (3) & 2.3 (2)   &  31 (4)  &  - \\
 2896.0 (4) & 0.9 (2)   &  14 (3)  &  - \\
\end{tabular}
\end{ruledtabular}
\end{footnotesize}
\end{center}
\end{table*}

\begin{figure*}
\centering
\includegraphics[width=\linewidth]{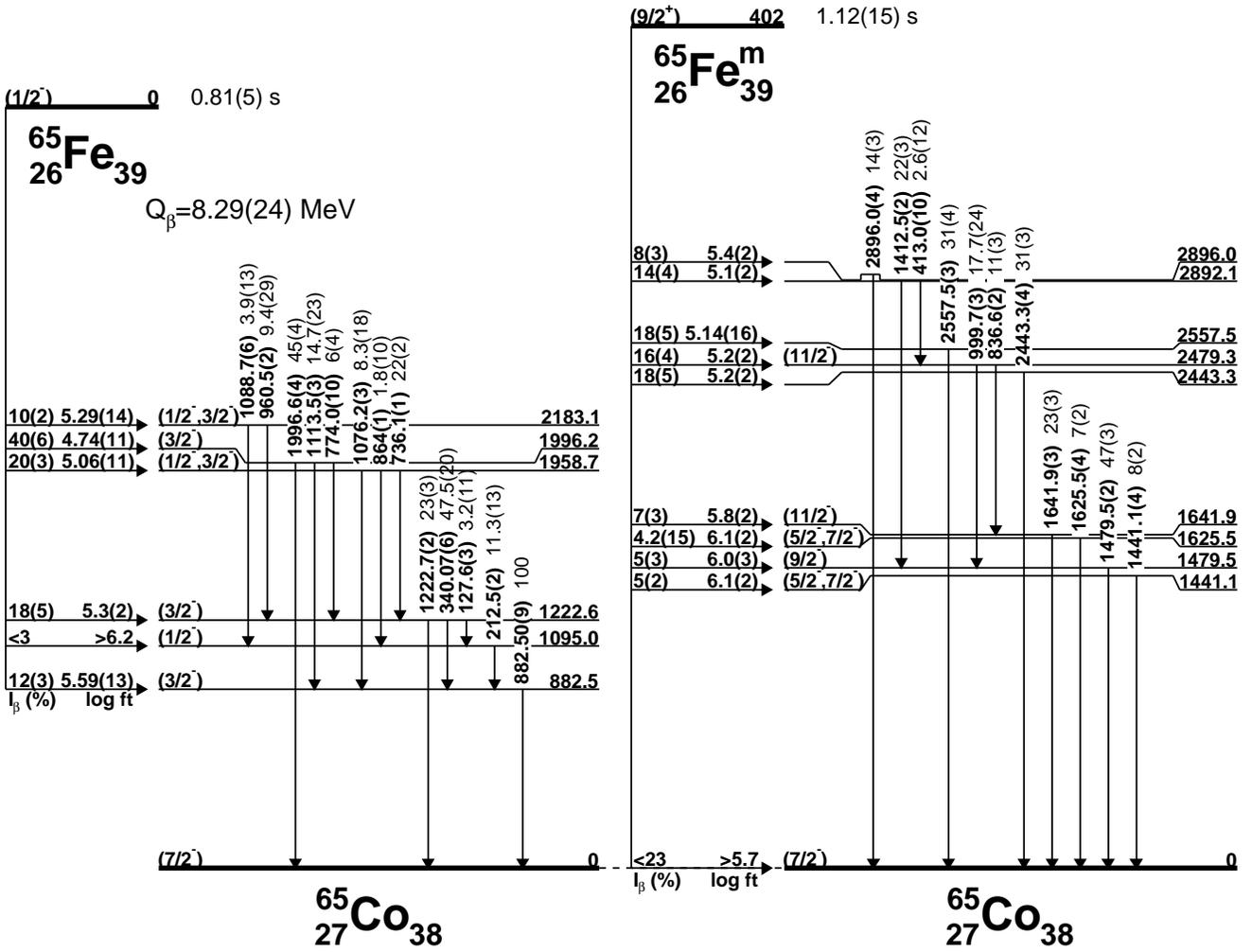}
 \caption{The two $^{65}$Fe decay schemes, see text for a detailed discussion.}
 \label{fig:65Fe_scheme}
\end{figure*}

Two independent level schemes can be constructed using the transitions and their coincidence relations given in Table \ref{tbl:65FelinesCo}. Starting with the most intense line in Table \ref{tbl:65FelinesCo}, the $883$-keV transition, a level at $883$ keV is deduced. Fig.~\ref{65Fe_b882g} shows the
$\gamma$ spectrum of prompt coincidence events with $\beta$-gated,
$883$-keV events. It reveals a strong coincidence with $340$-keV
$\gamma$ rays. In combination with the $1223$-keV cross-over transition,
the $1223$-keV level could be established as shown on the left
hand side in Fig.~\ref{fig:65Fe_scheme}. The coincident events at
$128$ and $213$~keV match perfectly the energy difference of
$340$~keV. Due to its higher intensity, the $213$-keV transition is
placed below the $128$-keV one and on top of the $883$-keV
$\gamma$ ray, establishing the $1095$-keV level. The $1959$-keV state could be established on the basis of the observed coincidences with
$1076$- and $736$-keV events, which differ by $340$~keV. The
$1076$-keV transition is thus placed on top of the $883$-keV level
and the $736$-keV $\gamma$ line on top of the $1223$-keV state. The
$864$-keV transition towards the $1095$-keV level was only observed
from coincidences with $213$-keV events (and one $883$-keV event),
see Table \ref{tbl:65FelinesCo}. The coincidence relationships between
$1114$- and $883$-keV events and a non-coincident cross-over
$1997$-keV line provide evidence for a $1996$-keV state. The
$774$-keV transition towards the $1223$-keV level is only observed
from coincidences with $340$-, $883$- and $1223$-keV events,
see Table \ref{tbl:65FelinesCo}. Based on the $961$-keV
coincidences with $340$- and $883$-keV events and coincidences
between the $1089$- and $213$-keV events, a level at $2183$-keV could be established, from which the $961$- and the $1089$-keV
transitions feed the $1223$-keV and $1095$-keV states,
respectively. The $961$-keV coincidences with $310$-keV $\gamma$ rays are due to the $963$-keV $^{65}$Ni component in this doublet. Based on all the arguments given above, the $\beta$-decay scheme is constructed as given on the left in Fig.~\ref{fig:65Fe_scheme}.

\begin{figure}
\centering
\includegraphics[width=0.82\linewidth]{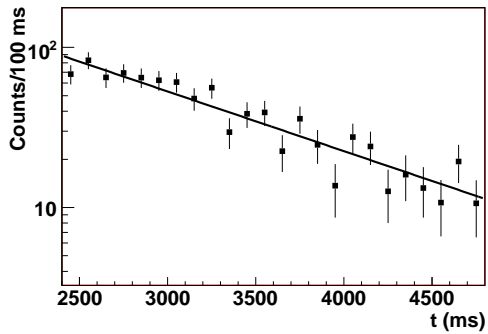}
 \caption{\small{Summed decay behavior with corresponding fit of the $\gamma$ lines at $340$, $736$, $883$ and $1997$ keV, representative of the $\beta$ decay of the $^{65}$Fe ground state. The line represents a single exponential fit resulting in a half-life value of $0.81(5)$ s.}}
 \label{65Fe_Decbeta883keV}
\end{figure}

\begin{figure}
\centering
\includegraphics[width=0.82\linewidth]{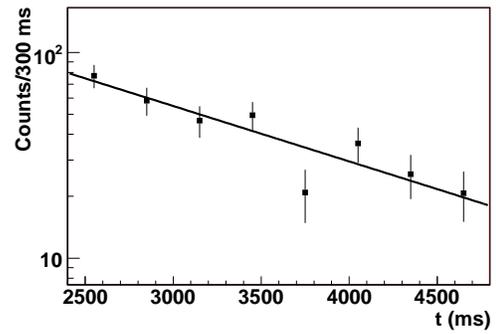}
\caption{\small{Summed decay behavior with corresponding fit of the $\gamma$ lines at $1000$, $1413$, $1480$ and $1642$ keV, representative of the $\beta$ decay of the $^{65m}$Fe isomer. The line represents a single exponential fit resulting in a half-life value of $1.12(15)$ s.}}
 \label{65Fe_Decbeta1000_1642keV}
\end{figure}

The $\beta$-decay scheme shown on the right of Fig.~\ref{fig:65Fe_scheme} is constructed following the same principles. The strongest transitions in this level scheme are the $1480$- and $1642$-keV transitions feeding the ground state. The $2479$-keV level could be established from the $1480$- and $1642$-keV coincidence relations with $1000$- and $837$-keV $\gamma$ rays, respectively. The $2892$-keV state could be established on the basis of coincidences between $1413$- and $1480$-keV events and observed $1000$-keV coincidences with $413$-keV events. The $\gamma$-ray transitions at $1441$, $1626$, $2443$, $2558$ and $2896$~keV, which do not show any $\gamma$ coincidences, are placed as ground-state transitions in this level structure based on spin and parity assignment considerations. These are discussed further below in this section.

By systematically shifting the relative excitation energy of both $^{65}$Fe $\beta$-decay paths, it was checked that the non-coincident lines at $1441$, $1626$, $2443$, $2558$ and $2896$~keV do not fit energy differences between states of the two structures simultaneously. As a result, two independent $^{65}$Fe $\beta$-decay paths without mutual coincident transitions have been deduced from data set I pointing to the presence of two $\beta$-decaying states in $^{65}$Fe. This is consistent with the recent discovery in Penning-trap mass measurements \cite{Blo_PRL_08} of a long-lived isomer (T$_{1/2} > 150$~ms) at a $402$-keV excitation energy in $^{65}$Fe, presumably of high spin ($9/2^{+}$), while the ground state has low spin ($1/2^{-}$) \cite{Blo_PRL_08_err}. Our gas cell is fast enough to allow detection of nuclei in the $100$-ms range. Two other excited levels are known at a lower energy in $^{65}$Fe; i.e., at $364$ and $397$~keV, respectively \cite{Dau_IAP_06}. In the event that the $402$-keV isomer does (partially) decay internally, a $\gamma$ ray has to be observed at $363$ or $402$~keV in the singles $\gamma$ spectrum, ignoring low-energy transitions. These transitions have not been observed in our $^{65}$Fe decay study. A proof of the presence of two $\beta$-decaying states in $^{65}$Fe would be the observation of $\gamma$ rays with different half-lives. The decay behavior of $\gamma$ rays belonging to the two different level schemes of Fig.~\ref{fig:65Fe_scheme} has been fitted by a single exponential function, shown in Figs.~\ref{65Fe_Decbeta883keV} and \ref{65Fe_Decbeta1000_1642keV}, resulting in half-life values of $1.12(15)$~s and $0.81 (5)$~s, respectively. Previously determined half-life values of $0.45 (15)$~s \cite{Cza_ZPA_94} and $1.3 (3)$~s \cite{Sor_NPA_00} were significantly different and Refs.~\cite{Sor_NPA_00,Blo_PRL_08} already suggested the presence of a $\beta$-decaying isomer in $^{65}$Fe to explain the discrepancy. However, the deduced half-life values are both longer than the value in Ref.~\cite{Cza_ZPA_94} and, therefore, cannot be explained on this basis.

The relative energy position of the two level structures could not be derived from the experimental data. For this reason, it is
a priori unclear whether both branches decay into a common ground
state of $^{65}$Co or if one of the branches decays into an isomeric state. In the latter case, the isomer has to reside at a relatively low excitation energy of $\sim 50$ keV or less, since the Penning-trap mass measurements of Ref.~\cite{Blo_PRL_08} did not resolve an isomeric state in $^{65}$Co. The relative $^{65}$Ni $\gamma$ intensities in data set I (lasers on iron) are found to be similar to those of data set II (lasers on cobalt). This is illustrated by the ratio of relative transition intensities $\frac{I_{rel}(II)}{I_{rel}(I)}$ from both data sets, see Table \ref{tbl:65FelinesNi}. Because more than $90 \%$ of the $\beta$ feeding goes directly to the $5/2^{-}$ ground state, it has also been checked that the absolute $\gamma$ intensity of the $310$-keV transition belonging to the $^{65}$Co decay is the same (within the uncertainties) in the direct $^{65}$Co decay (data set II) and via the $^{65}$Fe decay (data set I). The fact that these ratios and the absolute $\gamma$ intensity of the $310$-keV transition are similar further supports the assumption of the presence of only one $\beta$-decaying state in $^{65}$Co.

An intermediate conclusion here is that there is evidence for two excitation structures in $^{65}$Co without any interconnecting transitions, fed by two $\beta$-decaying states in $^{65}$Fe considerably differing in spin. Whether these level schemes are built on the same state and which one belongs to the decay of the high-spin (respectively low-spin) state needs further consideration. For the clarity of further discussion, we will attribute the left structure of Fig.~\ref{fig:65Fe_scheme} to the decay of the low-spin and the right structure to the high-spin state. This will be discussed in paragraph \ref{par:spin}.

\subsubsection{Deep-inelastic reaction at ANL}

The $\beta$-decay data alone are insufficient to claim the
(non-)existence of a $\beta$-decaying isomer in $^{65}$Co with absolute certainty. Crucial information on the $^{65}$Co level structure was extracted, however, from a deep-inelastic reaction study. In the present work, only the analysis of triple coincidence data (cubes) revealed useful results. As this type of reactions leads to the production of a wide range of projectile- and target-like fragments, the resulting complexity of the $\gamma$ spectra can be resolved by using coincidence techniques. However, in the case of weakly-produced nuclei such as $^{65}$Co, the analysis of $\gamma$-$\gamma$ coincidence events did not provide spectra sufficiently clean to allow for the  identification of new transitions and higher-order coincidences proved essential. Therefore, the success of the present analysis relied entirely on the high selectivity that could be achieved by analyzing triple gates in the coincidence cubes.


\begin{figure*}
\centering
\includegraphics[angle=-90,width=\linewidth]{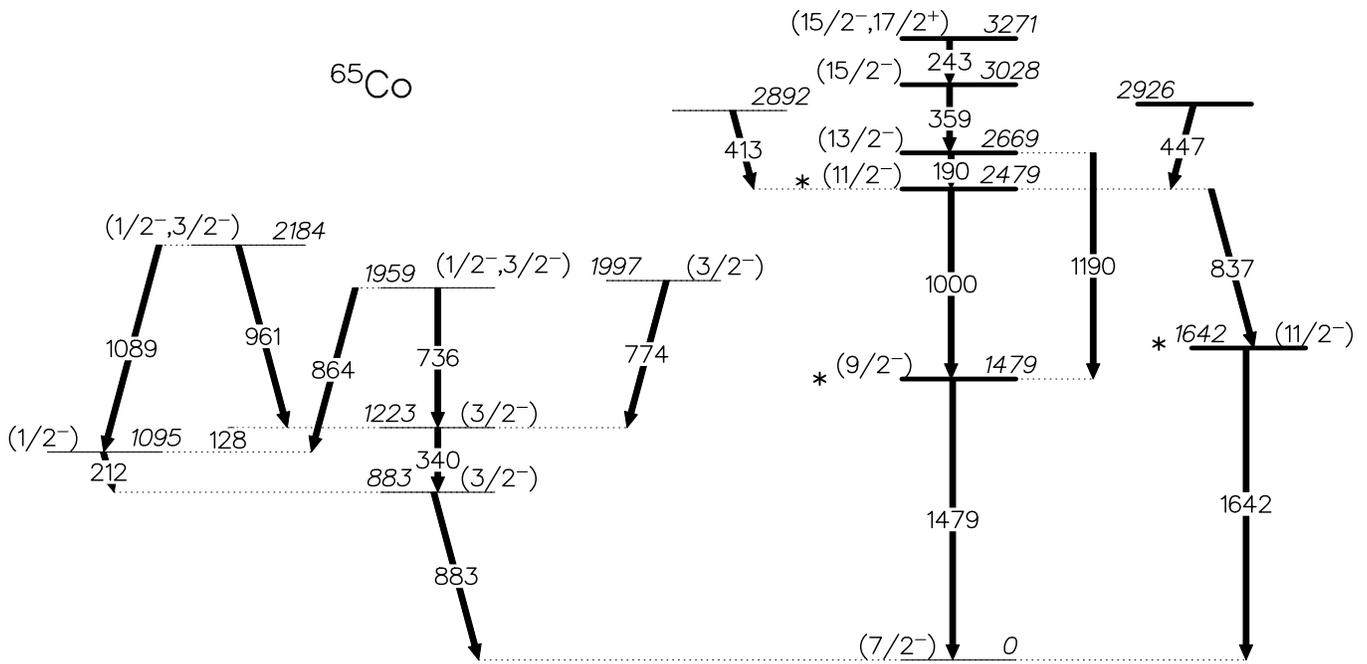}
 \caption{\small{$^{65}$Co level scheme deduced from the $^{64}$Ni+$^{238}$U deep-inelastic reaction by gating on any two coincident $\gamma$ rays observed in the $\beta$ decay of $^{65}$Fe. Levels drawn with thick lines were observed in the analysis of the PPP cube, those represented with thin lines were identified by placing double gates in the DDD cube. A star marks the states that were seen in both PPP and DDD cubes. All states seen in the DDD cube correspond to levels established in the $\beta$-decay data.}}
 \label{DI_65Co_scheme}
\end{figure*}

\begin{figure*}
\centering
\includegraphics[width=0.8\linewidth]{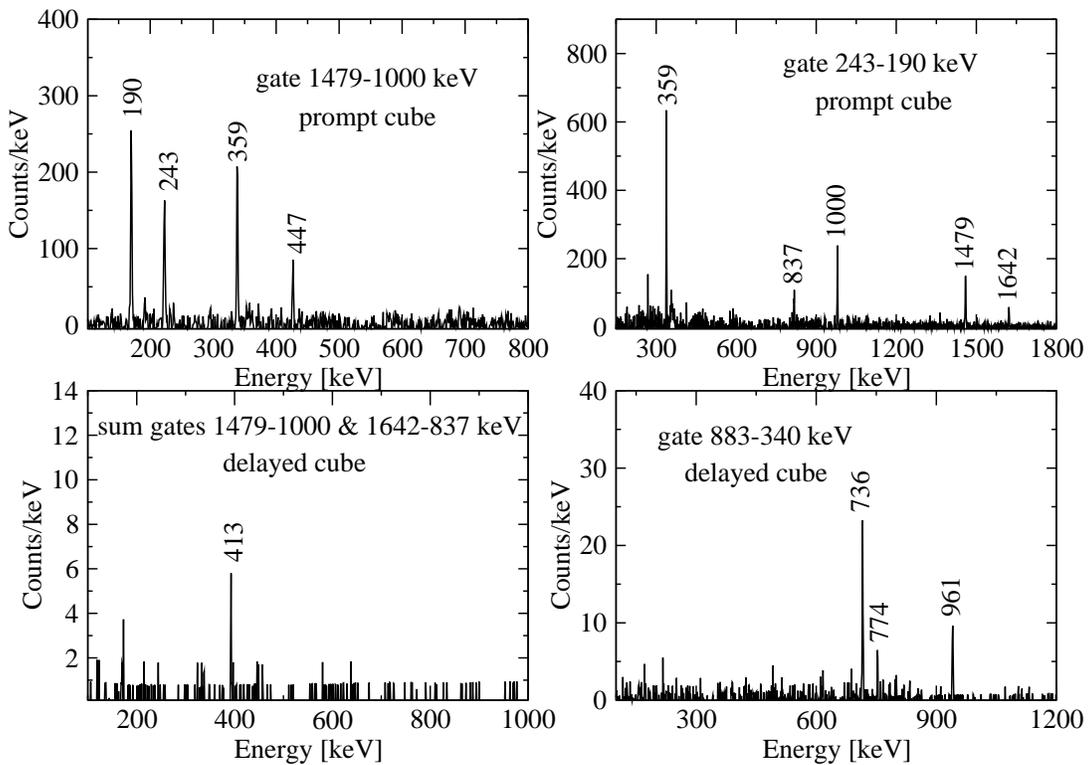}
 \caption{\small{Example of spectra from coincidence gates on transitions in $^{65}$Co in prompt (a and b) and delayed cubes (c and d) with the data taken in the $^{64}$Ni+$^{238}$U deep-inelastic reaction. Transitions in $^{65}$Co are identified by their energies.}}
 \label{d_gates}
\end{figure*}

Excited levels in $^{65}$Co were established by gating on any two coincident $\gamma$ rays observed in the $\beta$-decay work reported in the previous section. The Radware \cite{Rad_NPA_95} software package was used in the analysis. The study of triple coincidence data revealed the partial level scheme presented in Fig.~\ref{DI_65Co_scheme}. Examples of triple-coincidence spectra are found in Fig.~\ref{d_gates}. As seen in the spectra of Figs.~\ref{d_gates}c and \ref{d_gates}d, the delayed triple-coincidence spectra contain much less statistics when compared to the prompt coincidence data given in the upper part of the same figure (\ref{d_gates}a and \ref{d_gates}b). This is due to the fact that the observed delayed transitions result from the $\beta$ decay of $^{65}$Fe, which is only weakly produced in $^{64}$Ni+$^{238}$U deep-inelastic reactions. The investigation of the delayed cube is expected to confirm levels in $^{65}$Co reported in the previous section. On the other hand, the direct population of excited states can be studied by analyzing the prompt cube. In this case, the reaction mechanism favors the population of fairly high-spin yrast and near-yrast states, which might differ from those populated in $\beta$ decay.

Indeed, the analysis of the delayed cube confirmed most of the levels observed in the $\beta$-decay work of Fig.~\ref{fig:65Fe_scheme}. As already mentioned above, due to their high selectivity, only triple coincidences were used in the present study. Therefore, only cascades of three or more coincident $\gamma$ rays observed in the $\beta$-decay work could be confirmed by the investigation of the deep-inelastic data set. As a consequence, the $\gamma$ rays of 1076, 1114, 1223, and 1997 keV proposed to arise from the decay of the levels populated by the $(1/2^-)$ ground state of $^{65}$Fe, see Fig.~\ref{fig:65Fe_scheme}, and those of 1412, 1441, 1625, 2443, 2557 and 2896 keV believed to deexcite states populated in the decay of the  $(9/2^+)$ isomer could not be investigated. The levels proposed in the $\beta$-decay work and confirmed by the analysis of the DDD cube are drawn with thin lines or marked with a star in the level scheme of Fig.~\ref{DI_65Co_scheme}.

The prompt population of the levels in the left part of Fig.~\ref{DI_65Co_scheme} was studied by analyzing the PPP cube. No combination of gates set on transitions depopulating levels proposed (in the previous section) to receive feeding from the low-spin isomer in $^{65}$Fe provided useful information. This indicates a very weak cross-section for direct population in a deep-inelastic reaction and suggests low-spin values for these levels.

The prompt double-gates, set on the 1000-1479 keV and 837-1642 keV cascades corresponding to the two different decay paths of  the 2479-keV state, provided clear evidence for the presence of four low-energy coincident $\gamma$ rays with energies of 190, 243, 359, and 447 keV, see Fig.~\ref{d_gates}a. Based on the results of the analysis of all possible combinations of double gates on the observed transitions, the new $\gamma$ rays were arranged in the level scheme of $^{65}$Co and deexcite from the levels represented with thick lines in Fig.~\ref{DI_65Co_scheme}. The 413-keV transition, observed in the $\beta$-decay work, and also confirmed in the analysis of the delayed cube to be in coincidence with the 1000-1479 keV and 837-1642 keV intense transitions (see Fig.~\ref{d_gates}c), was not observed in the PPP cube. This indicates that the 2892-keV state does not belong to the yrast or near-yrast sequence in $^{65}$Co. The 2669-keV level identified in the prompt data was found to deexcite predominately via the 1190-keV transition (72\% branching) to the 1479-keV level.

For this type of deep-inelastic reaction, angular correlation measurements are commonly used to determine the multipolarity of $\gamma$ rays. However, in $^{65}$Co, two of the strongest prompt transitions that could possibly be used as gates, e.g., the 1480- and $1190$-keV $\gamma$ rays, form a doublet with two intense lines in $^{69}$Ga  \cite{Bak_PRC_82}, a nucleus strongly produced in the present reaction. This prevents reliable multipolarity assignments for the new states identified in $^{65}$Co in the deep-inelastic data set. Also, the statistics for the other observed transitions is too low to allow for an accurate angular-correlation analysis.

\subsubsection{Spin and parity considerations and assignments}\label{par:spin}

\begin{figure}
\centering
\includegraphics[width=\linewidth]{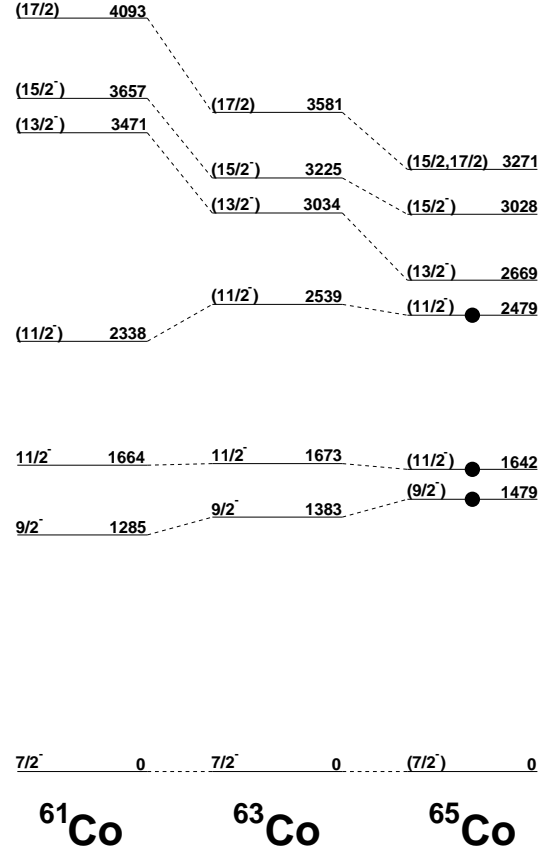}
 \caption{\small{Yrast and near-yrast levels in odd-mass $^{61-65}$Co that are systematically populated in deep-inelastic reactions. The $^{65}$Co levels at $1479$, $1642$, and $2479$~keV, respectively, are also observed in the $\beta$ decay of the $^{65}$Fe isomer (full circle). The levels of $^{61,63}$Co are taken from Ref.~\cite{Reg_PRC_96}.}}
 \label{DI_OddCoSystematics}
\end{figure}

Unfortunately, the deep-inelastic data did not allow spin and parity assignments to be made on the basis of angular correlations. Thus, for the levels observed in the analysis of the delayed data and marked with a star in Fig.~\ref{DI_65Co_scheme}, spin and parities assigned in the $\beta$-decay work are assumed, as will be discussed further in this paragraph. However, a comparison with the near-yrast levels of $^{59}$Co \cite{War_PRC_77} and $^{61,63}$Co \cite{Reg_PRC_96} populated in other deep-inelastic reactions, reveals a strikingly similar low-energy structure, as can be noticed in Fig.~\ref{DI_OddCoSystematics}. This suggests a $(7/2^{-})$ ground state, and $(9/2^{-})$, $(11/2^{-})$ and $(11/2^{-})$ levels at $1480$, $1642$ and $2479$~keV, respectively. Note, however, that the $2479$-keV level is also observed in the $^{65}$Fe$^{m}$ $\beta$-decay of the $(9/2^{+})$ isomer with a low log~$ft$ value of $5.2 (2)$. This value is inconsistent with the proposed negative parity. One should, nevertheless, also realize that the reported log~$ft$ values are essentially lower limits due to the possibility of unobserved $\gamma$-ray activity from high-energy levels. Because of the systematic similarity with the $^{59-63}$Co level structures obtained in deep-inelastic reactions, the $(11/2^{-})$ assignment to the $2479$-keV level is suggested. The states on top of this $2479$-keV state most likely form the $(13/2^{-})$, $(15/2^{-})$ sequence, similarly to $^{61,63}$Co \cite{Reg_PRC_96} as can be seen in Fig.~\ref{DI_OddCoSystematics}, but positive parity cannot be entirely disregarded. It is worth mentioning that 15/2 is the maximum spin value of negative parity that can be achieved within the $\pi f_{7/2}^{-1}\nu(p_{3/2}f_{5/2}p_{1/2})^{-2}$ configuration. The highest observed level at 3271 keV might have either $J^\pi$=17/2$^+$ or 15/2$^-$.

In the deep-inelastic data there is also no linking transition observed between the low-spin levels (left part of Fig.~\ref{DI_65Co_scheme}) and the high-spin levels (right part of Fig.~\ref{DI_65Co_scheme}). They are placed, as in Fig.~\ref{fig:65Fe_scheme}, on the $(7/2^{-})$ ground state of $^{65}$Co, as the only argument for a low-spin isomer in $^{65}$Co, the feeding of the previously $(1/2^{-})$ assigned $1274$-keV level in $^{65}$Ni \cite{Fly_PRL_78}, has been argued against in paragraph \ref{par:65Ni}.

In Fig.~\ref{fig:65Fe_scheme}, all the states fed by the $(1/2^{-})$
$^{65}$Fe ground state (log $ft \le 5.6(2)$), and subsequently
decaying towards the $(7/2^{-})$ ground state, can be
assigned as $(3/2^{-})$ levels; i.e., the states at $883$, $1223$ and $1996$~keV. Allowed Gamow-Teller transitions and $\gamma$ transitions with multipolarity less than three are assumed. The states at $1959$ and $2183$~keV are also significantly fed in $\beta$ decay from the
$(1/2^{-})$ ground state, but their decay towards the $(7/2^{-})$ $^{65}$Co ground state is not observed. It is tempting to assign these states a spin and parity of $1/2^{-}$, but from Weisskopf estimates, a $3/2^{-}$ assignment cannot be ruled out. The state at $1095$~keV is fed by low-spin levels and decays preferentially into the $(3/2^{-})$ state at $883$~keV rather than towards the $(7/2^{-})$ ground state. Hence, a $(1/2^{-})$ assignment is tempting, but again $(3/2^{-})$ cannot be disregarded. Note also that the $1095$-keV level is not fed in $\beta$ decay, pointing to the fact that its structure is significantly different from the other observed low-spin states; see also section \ref{sec:65Co_Interpretation} for more details.


\subsection{$^{67}$Fe decay to $^{67}$Co}

\begin{figure*}
\centering
\includegraphics[width=\linewidth]{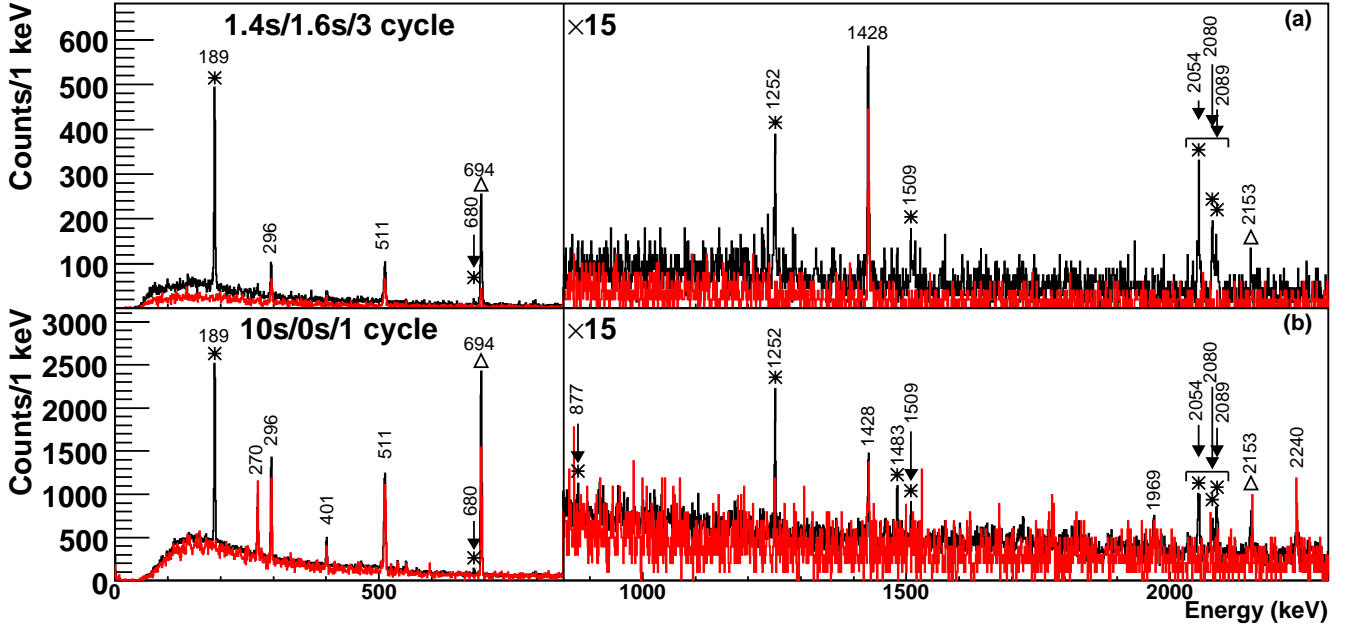}
 \caption{\small{The vetoed, $\beta$-$\gamma$ spectrum with the lasers tuned to ionize iron in black and with the lasers off normalized to the laser-on time in red (on-line version only) in the $1.4$s/$1.6$s/$3$ cycle (top, data sets IV and V, respectively) and the $10$s/$0$s/$1$ cycle (bottom, data sets VI and VII, respectively). The lines from $^{67}$Fe $\beta$ decay are marked with a star and those from $^{67}$Co with an open triangle. Contaminant lines are not marked, see in the text for further information.}}
 \label{fig:67Fe_spectrum}
\end{figure*}

Fig.~\ref{fig:67Fe_spectrum}(a) presents the vetoed, $\beta$-gated
$\gamma$ spectra from data set IV with the lasers tuned resonantly to iron in black and from data set V with the lasers off normalized to
the laser-on time in red. Fig.~\ref{fig:67Fe_spectrum}(b) provides the vetoed, $\beta$-gated $\gamma$ spectra from data set VI with the
lasers tuned resonantly to iron in black and from data set VII with
the lasers off normalized to the laser-on time in red. The full
circles indicate lines from $^{67}$Fe, open squares from $^{67}$Co
and open triangles from contaminant $\beta$ decay. From data set IV
the $^{67}$Fe half-life was extracted with a single exponential fit
of the $189$-keV time dependence during the decay period
($T_{1/2}=416(29)$~ms) \cite{Pau_PRC_08}. From data set VI the
$^{67}$Co level scheme was constructed \cite{Pau_PRC_08}.

\begin{table*}[htb]
\begin{center}
\caption{$^{67}$Co transitions from data set IV are indicated by
their energy $E$~(keV), the corresponding off resonant subtracted
peak count rate A$_{\gamma}$ corrected for $\beta$ efficiency and
$\gamma$ intensity I$_{rel}$ relative to the $189$-keV transition
($100 \ \%$). The last six lines give upper $\gamma$-intensity
limits ($95 \ \%$ confidence limit) of important transitions.
Multiply $I_{rel}$ by 0.847 (36) to get absolute
$\gamma$ intensities. The $\gamma$ energies of coincident events are listed in the last column with the number of observed $\beta$-$\gamma$-$\gamma$ coincidences between brackets.\label{tbl:67Colines}}
\begin{footnotesize}
\begin{ruledtabular}
\begin{tabular}{ccccc}
 E (keV)    & A$_{\gamma}$ (cts/h)
                                        & I$_{rel}$ (\%)
                                            & Coincident $\beta\gamma$-events \\
\hline
 188.93 (8) & 186 (6)   & 100     & 571(22), 877(7), 1179(10), 1483(5), 2054(34), 2080(10), 2089(29) \\
 491.55 (11)& 99 (15)   & 99 (29) & - \\
 571.4 (2)  & 3.4 (10)  & 3.7 (14) & 189(17), 1483(1), 1509(3) \\
 680.4 (2)  & 6.6 (11)  & 8 (3)   & - \\
 876.9 (6)  & 3.4 (10)  & 5 (2)   & 189(9), 1179(5), 1368(2) \\
 1178.9 (3) & 1.9 (6)   & 3.3 (14) & 189(7), 877(5)\\
 1251.9 (4) & 5.4 (19)  & 10 (4)  & 1483(6), 1509(4)\\
 1368.0 (10)& 0.5 (4)   & 1.1 (9)\footnote{The relative intensity is based on the number of coincidences with $\beta$-gated $877$-keV $\gamma$ rays.} & 877(2) \\
 1483.2 (2) & 3.4 (7)   & 7 (2)   & 189(5), 571(1), 1252(6) \\
 1508.9 (3) & 3.8 (8)   & 8 (3)   & 189(1), 571(3), 1252(4)\\
 2054.2 (2) & 6.9 (9)   & 18 (5)  & 189(35) \\
 2079.8 (5) & 3.3 (9)   & 8 (3)   & 189(14) \\
 2088.7 (2) & 6.3 (9)   & 16 (5)  & 189(29) \\
\hline
 2243.3 (2) & -         & $<10$   & - \\
 2269.0 (3) & -         & $<3.9$  & - \\
 2277.6 (2) & -         & $<3.3$  & - \\
 2734.8 (2) & -         & $<4.5$  & - \\
 2760.6 (3) & -         & $<4.4$  & - \\
 2769.2 (2) & -         & $<0.9$  & - \\
\end{tabular}
\end{ruledtabular}
\end{footnotesize}
\end{center}
\end{table*}

Lines present in the spectra with the lasers on and absent with the
lasers off can unambiguously be identified as coming from $^{67}$Fe
$\beta$ decay. The isomeric behavior of a laser-enhanced line at
$492$~keV was already evidenced and extensively discussed in
Ref.~\cite{Pau_PRC_08}. The most intense transitions in the
$^{67}$Fe decay have energies of $189$, $492$, $680$,
$1252$, $1483$, $1509$, $2054$, $2080$ and $2089$~keV. The strong $\gamma$ line at $694$~keV originates from
$^{67}$Co $\beta$ decay \cite{Wei_PRC_99}. The contaminant lines
are originating from doubly-charged molecules. The lines at
$270$, $1969$ and $2240$~keV are from the decay of
$^{106}$Tc; the lines at $296$ and $401$~keV from $^{102}$Nb;
and the line at $1428$~keV from $^{94}$Sr. Molecules are formed
with CO, O$_{2}$ and $^{40}$Ar, respectively.

\begin{figure*}
\centering
\includegraphics[width=\linewidth]{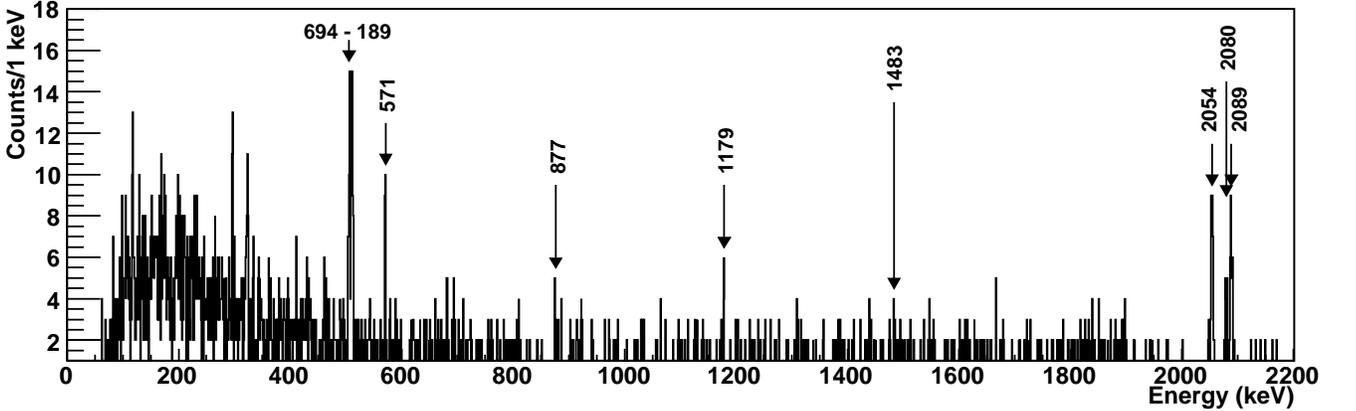}
 \caption{\small{The $\gamma$ spectrum of prompt coincidences with $\beta$-gated $189$-keV events (from data set VI). The structure indicated by $694-189$ is due to Compton scattering of the $694$-keV line.}}
 \label{67Fe_b189g}
\end{figure*}

\begin{figure}
\centering
\includegraphics[width=\linewidth]{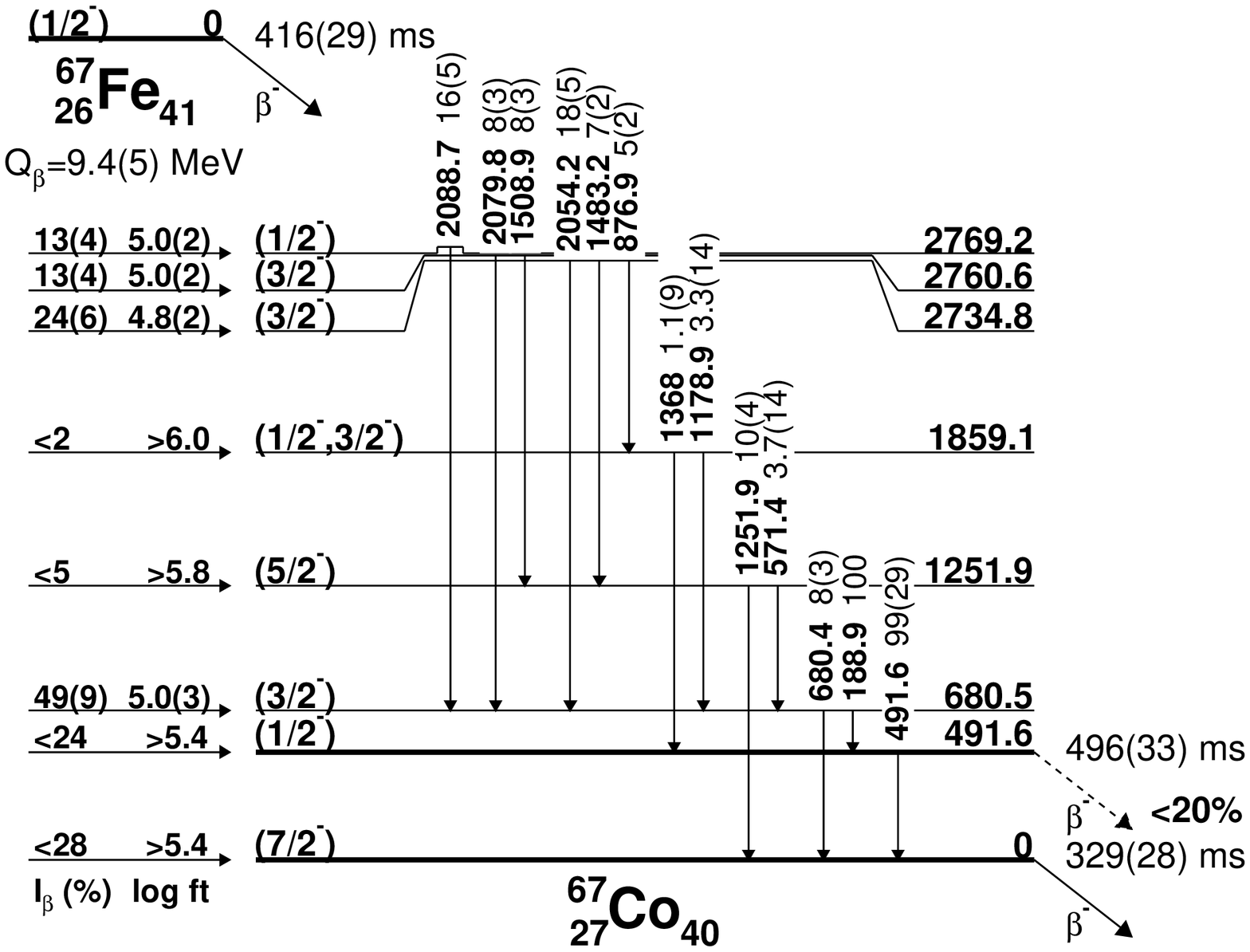}
 \caption{Partial $^{67}$Co level scheme. Multiply the relative $\gamma$ intensities by $0.85 (4)$ to obtain absolute $\gamma$ strengths.}
 \label{fig:67Co_scheme}
\end{figure}

The $492$-keV and $680$-keV levels have already been
established from correlations measured in the seconds time range
\cite{Pau_PRC_08}. However, the construction of the higher-energy
structure of $^{67}$Co, given in Fig.~\ref{fig:67Co_scheme}, has
not been discussed in detail yet. A solid basis is given by the
$\gamma$ spectrum of coincident $189$-keV events, shown in
Fig.~\ref{67Fe_b189g}. The coincident $571$-keV line and a
$1252$-keV cross-over $\gamma$ ray give evidence for the state at
$1252$~keV. The $1859$-keV state is placed, based on the $1179$-keV
line in the spectrum of Fig.~\ref{fig:67Co_scheme} together with the, albeit weak, $1368$-keV cross-over transition. Evidence for a $2735$-keV level is delivered by the $189$-keV $\gamma$ ray being coincident with the $2054$-keV line and the presence of
the $877$-keV and $1483$-keV events, which are coincident with
$1179$-keV and $1252$-keV events, respectively. The $2761$-keV
state could be established from the $189$-keV $\gamma$ ray being coincident with the $2080$-keV line and the $1509$-keV events being coincident with $1252$-keV events.
The level at $2769$~keV, finally, is placed on the basis of the
$2089$-keV line in the spectrum of Fig.~\ref{67Fe_b189g}. The spectrum also indicates an intense peak at $505$ keV (marked with ''$694-189$''), which contains the $189$-keV coincident Compton events originating from the $694$-keV transition.

The $\gamma$ intensities indicated in Fig.~\ref{fig:67Co_scheme}
are relative to the $189$-keV transition ($I_{\gamma}=100$). The
assigned transitions, count rates and
relative intensities are summarized in Table \ref{tbl:67Colines}.
The data do not leave room for $\beta$ feeding towards the
$^{67}$Co ground state by comparing the total $\gamma$ intensity of
$^{67}$Co and $^{67}$Ni. The $\beta$ feeding towards the excited
$^{67}$Co states is based on missing $\gamma$ intensities. As can
be seen in Fig.~\ref{fig:67Co_scheme}, the total feeding is shared
$50-50$ between the state at $680$~keV and the group of levels at
$2769$, $2761$ and $2735$~keV. It is remarkable that the
latter group of states decays to the level at $680$~keV, but not
towards the ground state nor the state at $492$~keV. Therefore,
Table \ref{tbl:67Colines} also provides upper limits ($95 \%$
confidence limit) for the respective cross-over transitions.
Compared to the $2054$-keV transition, all the cross-over
transitions, except for the $2243$-keV $\gamma$ ray, are at least $4$ times weaker. The larger upper limit for the $2243$-keV transition is due to the presence of a $^{106}$Tc contamination at
$2240$~keV. The transitions at $571$, $877$, $1179$ and
$1368$~keV were firmly assigned to the $^{67}$Fe $\beta$ decay following the inspection of the observed $\beta$-$\gamma$-$\gamma$ coincidences. The isomeric
$492$-keV state has been fully characterized by correlations with
$\beta$-gated $189$- and $694$-keV $\gamma$ events.

The spins and parities of the four lowest $^{67}$Co levels are
assigned in Ref.~\cite{Pau_PRC_08}. The $^{67}$Fe ground state was
initially assigned a spin and parity of $J^{\pi}=(5/2^{+})$ from
isomeric $^{67m}$Fe $\gamma$ decay \cite{Saw_EPJ_03}, but this
would be inconsistent with the strong $\beta$ branch towards the
$(3/2^{-})$ state at $680$~keV. On the other hand, a $(1/2^{-})$
assignment for the $^{67}$Fe ground state, as proposed in
Ref.~\cite{Blo_PRL_08}, is in agreement with the proposed level
scheme. It also explains the lack of feeding towards both the $(7/2^{-})$ ground state and the $1252$-keV $(5/2^{-})$ level. The low log $ft$ limit of $5.4$ for the $\beta$ feeding of the $(1/2^{-})$ isomer will be discussed in next section. The strong $\beta$ decay towards the states at $2735$ (log $ft=4.8(2)$), $2761$ (log $ft=5.0(2)$) and $2769$~keV (log $ft=5.0(2)$) gives strong support for spin and parity  of $(1/2^{-})$ or $(3/2^{-})$.

The $2735$- and $2761$-keV levels are observed to decay to the $(5/2^{-})$ state at $1252$~keV with an intensity similar to that feeding the $(3/2^{-})$, $680$-keV level, while the $2769$-keV level is only observed to decay to the $680$-keV state and not to the $1252$-keV level. From Weisskopf estimates, E$2$ and M$1$ transitions from the $~2.75$-MeV levels towards the $1252$-keV state are, respectively, three orders of magnitude and two times less probable than M$1$ transitions to the $680$-keV level. Hence, the $2735$- and $2761$-keV states can tentatively be assigned a $(3/2^{-})$ spin and parity and the $2769$-keV level $(1/2^{-})$ quantum numbers. The state at $1859$~keV is observed to decay to both the $(3/2^{-})$ state at $680$~keV and the $(1/2^{-})$ level at $492$~keV and not to the $7/2^{-}$ ground state. This restricts its spin and parity to $1/2^{-}$ or $3/2^{-}$, despite the lack of direct $\beta$ feeding from the $^{67}$Fe ground state.


\section{Interpretation}

\subsection{Levels in $^{65}$Ni}\label{sec:65Ni_Interpretation}

The $^{65}$Co ground state has spin and parity $(7/2^{-})$, which arises from the $\pi f^{-1}_{7/2}$ proton-hole. The dominant $\beta$-decay path is the Gamow-Teller conversion of a $f_{5/2}$-neutron into a $f_{7/2}$-proton feeding states in $^{65}$Ni, which can be interpreted as a $\nu f^{-1}_{5/2}$ neutron-hole coupled to $^{66}$Ni. By far, the strongest $\beta$ strength is observed toward the $5/2^{-}$ ground state, which can be interpreted rather naturally as the $\nu f^{-1}_{5/2}$ neutron-hole state.

The other levels that are fed in $\beta$ decay reside at an excitation energy of $1141$ and $1274$~keV, which are tentatively assigned $(7/2^{-},9/2^{-})$ and $(5/2^{-})$ spin and parity, respectively. The low-energy level structure of $^{65}$Ni can then be interpreted as a $\nu f^{-1}_{5/2}$, $\nu p^{-1}_{1/2}$ or $\nu p^{-1}_{3/2}$ neutron-hole coupled to the $^{66}$Ni level structure. The $1/2^{-}$ state at $63$~keV is the $\nu p^{-1}_{1/2}$ neutron-hole state and the $3/2^{-}$ level at $310$~keV is the $\nu p^{-1}_{3/2}$ neutron-hole state. The $1274$-keV and the $1141$-keV levels can be interpreted as the coupling of the $\nu f^{-1}_{5/2}$ hole to the $2^{+}$ state in $^{66}$Ni, which lies at $1426$ keV (see Fig.~\ref{fig:65Co_int}).

\subsection{Levels in $^{65}$Co}\label{sec:65Co_Interpretation}

\begin{figure}
\centering
\includegraphics[width=\linewidth]{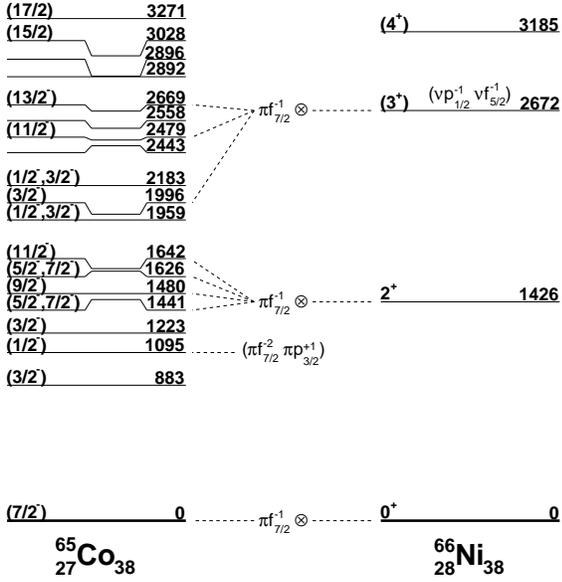}
 \caption{The $^{65}$Co levels interpreted as proton-hole states coupled to $^{66}$Ni coexisting with a proton intruder state at $1095$~keV.}
 \label{fig:65Co_int}
\end{figure}

As will be discussed in this paragraph, $^{65}$Co can be interpreted in terms of two coexisting structures. On the one hand, states are suggested arising from a $\pi f^{-1}_{7/2}$ proton hole coupled to the first excited $2_{1}^{+}$ and $3_{1}^{+}$ states of the $^{66}$Ni core, see Fig.~\ref{fig:65Co_int}. On the other hand, the $(1/2^{-})$ state at $1095$ keV is suggested to arise from proton excitations across the $Z=28$ shell closure. This assignment is based on the similarity with the established $(1/2^{-})$ proton intruder state in $^{67}$Co, which is observed in the $^{67}$Fe $\beta$ decay \cite{Pau_PRC_08}. In addition, the comparison with the $^{67}$Co structure indicates that the $(3/2^{-})$ state at $1223$ keV can be interpreted as the first member of the rotational band built on top of the $(1/2^{-})$ proton intruder state.

The most straightforward configuration of the suggested $(1/2^{-})$ $^{65}$Fe ground state \cite{Blo_PRL_08} is $\pi f^{-2}_{7/2} \nu p^{-1}_{1/2}$. This assignment is, indeed, consistent with the ground state of the $^{66}$Co isotone, which is proposed to be the $3^{+}$ member of the $\pi f^{-1}_{7/2} \nu p^{-1}_{1/2}$ configuration \cite{Mue_PRC_00,Iva_The_07}. For both isotones, the preferred decay mode is a Gamow-Teller conversion of a $f_{5/2}$-neutron into a $f_{7/2}$-proton \cite{Mue_PRC_00}. Since the dominant neutron configuration is identical in the two isotones, a similar $\beta$-decay pattern is expected for both cases. In the $^{66}$Co $\beta$ decay, two levels at $2672$ and $3228$~keV with a dominant $\nu p^{-1}_{1/2} \nu f^{-1}_{5/2}$ configuration, coupling to a respective spin and parity of $2^{+}$ or $3^{+}$, are strongly fed \cite{Mue_PRC_00}. In the $^{65}$Fe $\beta$ decay, strong allowed Gamow-Teller decay is expected to $J^{\pi}=1/2^{-}$ and/or $3/2^{-}$ states with a dominant $\pi f^{-1}_{7/2} \nu p^{-1}_{1/2} \nu f^{-1}_{5/2}$ configuration, which is a $\pi f^{-1}_{7/2}$ proton-hole coupled to the strongly fed $2^{+}$ or $3^{+}$ states of $^{66}$Ni.

The strongest feeding is observed to the $(3/2^{-})$ level at $1996$~keV with a log~$ft$ value of $4.74 (11)$. This is very similar to the log~$ft$ value of $4.2 (5)$ of the $^{66}$Co $\beta$ decay towards the $3^{+}$ state at $2670$~keV \cite{Mue_PRC_00}. On this basis, the $1996$-keV state in $^{65}$Co can be interpreted as arising from a coupling of the $\pi f^{-1}_{7/2}$ proton hole with the $3^{+}$ state, and a large percentage of the $\pi f^{-1}_{7/2} \nu p^{-1}_{1/2} \nu f^{-1}_{5/2}$ configuration is assigned, as depicted in Fig.~\ref{fig:65Co_int}. One of the states at either $1959$ or $2183$~keV may well be the $1/2^{-}$ member of the multiplet, but, in this case, the higher log~$ft$ values indicate that the proposed configuration is less dominant than in the $3/2^{-}$ member.

The high-spin members of the $\pi f^{-1}_{7/2} \nu p^{-1}_{1/2} \nu f^{-1}_{5/2}$ multiplet are not fed in the $\beta$ decay of the $(1/2^{-})$ ground state and should rather be searched for in the decay of the $9/2^{+}$, $^{65}$Fe$^{m}$ isomer. The $(9/2^{+})$ $^{65}$Fe isomeric state, however, has to involve the $\nu g^{+1}_{9/2}$ configuration and, similar to the $(1/2^{-})$ ground state, the Gamow-Teller conversion of a $f_{5/2}$-neutron into a $f_{7/2}$-proton is expected to be the preferred decay mode. The $2479$-keV state is, despite its apparent low log~$ft$ value, tentatively assigned $(11/2^{-})$ and the $2669$-keV level tentatively $(13/2^{-})$, which would be the highest possible spin formed by a $\pi f^{-1}_{7/2} \otimes 3^{+}$ coupling. There is, however, no further evidence to support such an interpretation.

The states at $1480$ and $1642$~keV were assigned $(9/2^{-})$ and
$(11/2^{-})$, respectively, from the systematics of $^{59,61,63}$Co
structures revealed in deep-inelastic experiments
\cite{War_PRC_77,Reg_PRC_96}. Since the configuration of the
$(9/2^{-})$ and $(11/2^{-})$ states in the lighter cobalt
isotopes were identified as corresponding to a coupling of $\pi f^{-1}_{7/2}$ with the first excited $2^{+}_{1}$ state \cite{Ste_PRC_71}, it is tempting to assign the same configuration to the $1480$- and $1642$-keV states. The $2_{1}^{+}$ excitation energy in $^{66}$Ni of $1425$~keV and the observed weak $\beta$ decay from the $(9/2^{+})$ isomer to these levels is consistent with this interpretation. Two other states, at $1441$ and $1626$~keV, exhibit similar excitation energies and $\beta$-decay strengths, see Fig.~\ref{fig:65Fe_scheme}, which indicate that these might be other members of the multiplet. A first-forbidden decay from the $(9/2^{+})$ level can only feed levels with $J \ge 5/2$. Moreover, the $1441$- and $1626$-keV states are only observed to decay to the $(7/2^{-})$ ground state. Hence, they are good candidates for the $5/2^{-}$ and $7/2^{-}$ members of the multiplet. The only missing member of the multiplet, at this point, is the $3/2^{-}$ level, but with the $(3/2^{-})$ levels at $883$ and $1223$~keV, there are two possible candidates in this case as well.

Both $(3/2^{-})$ states are rather strongly fed by direct $\beta$ decay, but the $(1/2^{-},3/2^{-})$ level at $1095$~keV gets almost no $\beta$ feeding (log~$ft > 6.2$) and, therefore, must be associated with a fundamentally different configuration. This low-energy structure is very similar to the one seen in $^{67}$Co, where the $\beta$ decay is strongly hindered to a $(1/2^{-})$ state at $492$~keV, which was interpreted as a $\pi(1p-2h)$ intruder state \cite{Pau_PRC_08}. Analogously, the $1095$-keV level is most likely a deformed $(1/2^{-})$ state involving proton excitations across $Z=28$. As in $^{67}$Co, the intruder state appears low in energy because of strong attractive proton-neutron residual interactions \cite{Hey_PR_83}, that are maximized by the weakening of the $N=40$ subshell closure.

From this point of view, the first excited $(3/2^{-})$ level can be interpreted as the $\pi f^{-1}_{7/2}$ coupling to the $2_{1}^{+}$ state, and the second excited $(3/2^{-})$ level as the first rotational band member of the $(1/2^{-})$ proton intruder, in analogy with $^{67}$Co. Nevertheless, both states are significantly fed in $^{65}$Fe $\beta$ decay with respective log~$ft$ values of $5.59 (13)$ and $5.3 (2)$. Their similar strengths indicate significant configuration mixing for both levels. With the $883$-, $1441$-, $1480$-, $1626$-, and $1642$-keV states interpreted as members of the $\pi f^{-1}_{7/2} \otimes 2^{+}$ quintet, coexisting with the $1095$- and $1223$-keV states interpreted as proton intruder configurations, we describe all levels below $1.7$ MeV.


\subsection{Levels in $^{67}$Co}\label{sec:67Co_Interpretation}

The $^{67}$Fe ground state requires, contrary to $^{65}$Fe,
excitations across $N=40$ to account for its $(1/2^{-})$ spin and
parity. In fact, the ground state of the $^{68}$Co isotone has been
interpreted as associated with a $\pi f^{-1}_{7/2} \nu g^{+2}_{9/2} \nu p^{-1}_{1/2}$ configuration \cite{Mue_PRC_00}. Hence, the ground state of $^{67}$Fe consists most likely of a dominant $\pi f^{-2}_{7/2} \nu g^{+2}_{9/2} \nu p^{-1}_{1/2}$ configuration and, as in the $A=65$ case discussed in the previous paragraph, one should expect a similar $^{68}$Co and $^{67}$Fe $\beta$-decay pattern of a $\nu f_{5/2} \rightarrow \pi f_{7/2}$ allowed Gamow-Teller conversion. However, the $^{68}$Co $\beta$-decay study revealed that there is no level in $^{68}$Ni with a dominant $\nu g^{+2}_{9/2} \nu f^{-1}_{5/2} \nu p^{-1}_{1/2}$ configuration in its wave function. In $^{68}$Ni, two levels at $4027$ and $4165$~keV were interpreted as arising from this multiplet, but with considerable mixing with the $\pi f^{-1}_{7/2} \pi p^{+1}_{3/2}$ configuration. Therefore, the $\beta$-decay pattern of $^{67}$Fe and $^{68}$Co may not be as strikingly similar as in the case of $^{65}$Fe and $^{66}$Co.

Strong feeding is observed to three levels at $\sim 2.75$~MeV with low log~$ft$ values of $\sim 5.0$. This compares well with the $\beta$ decay of the $^{65}$Fe ground state to three levels at $\sim 2$~MeV, and is indicative of their structure consisting of a dominant $\pi f^{-1}_{7/2} \nu g^{+2}_{9/2} \nu f^{-1}_{5/2} \nu p^{-1}_{1/2}$ configuration. However, the $\gamma$ decay in $^{65}$Co and $^{67}$Co is substantially different. In Figs.~\ref{fig:65Fe_scheme} and \ref{fig:67Fe_spectrum} it can be noticed, for instance, that the $1996$-keV level in $^{65}$Co strongly decays towards the ground state, while no ground state transitions have been observed from any of the $\sim 2.75$-MeV levels in $^{67}$Co. Thus, there appears to also be a significant structural change when going from the $N=38$ nucleus, $^{65}$Co, to the $N=40$ nucleus, $^{67}$Co. The low-energy states at $680$ and $1252$~keV have been discussed already in previous work \cite{Pau_PRC_08} and were interpreted as the first members of a rotational band built on the $(1/2^{-})$ proton intruder state.


\section{Conclusions}

The mass $A=65$ $\beta$-decay chain from iron down to nickel and
the $^{67}$Fe $\beta$ decay have been investigated at the LISOL
facility. The $^{65}$Co structure has been studied in detail from the combined analysis of $\beta$ decay and complementary deep-inelastic data, which were taken at ANL. A new $^{65}$Fe and a more detailed $^{67}$Fe decay scheme have been presented and discussed.

The $\beta$ decay of $^{65}$Fe is feeding two independent
level structures originating from a $(1/2^{-})$ ground state and a
$(9/2^{+})$ isomeric state. The half-lives of both states have been determined as $T_{1/2}=0.81(5)$~s and $T_{1/2}=1.02(14)$~s, respectively. The deduced $^{65}$Co structure can be interpreted as arising from the coupling of a $\pi f^{-1}_{7/2}$ proton-hole state with core levels of $^{66}$Ni, coexisting with a proton intruder state at $1095$~keV.

The subsequent $\beta$ decay of $^{65}$Co is revisited in the present work. Apart from the wrongly assigned $883$- and $340$-keV transitions, which are now unambiguously placed in the $^{65}$Fe decay scheme, the $^{65}$Co decay scheme is found to be consistent with that proposed in Ref.~\cite{Bos_NPA_88}. The two independent level structures observed in the $^{65}$Fe $\beta$ decay are placed on one common $(7/2^{-})$ ground state. As a consequence, we claim that the previous $1/2^{-}$ assignment of the $1274$-keV level in $^{65}$Ni \cite{Fly_PRL_78} is incorrect. Instead, a $(5/2^{-})$ spin and parity is suggested. The observed low-energy structure of $^{65}$Ni is interpreted as arising from $\nu f^{-1}_{5/2}$, $\nu p^{-1}_{1/2}$ and $\nu p^{-1}_{3/2}$ neutron-hole configurations coupled to the $^{66}$Ni core structure.

The $^{65}$Fe and $^{67}$Fe ground states feature
similar $\beta$-decay patterns. Strong feeding is observed to three
high-energy $(1/2^{-}),(3/2^{-})$ nickel core-coupled states
arising from $\nu p^{-1}_{1/2} f^{-1}_{5/2}$ and $\nu g^{+2}_{9/2}
p^{-1}_{1/2} f^{-1}_{5/2}$ neutron configurations, respectively,
and to a $(3/2^{-})$ level suggested to arise from a proton
intruder configuration, which in the case of $^{65}$Co is strongly mixed with the $2^{+}$ core-coupled configuration. However, there is also evidence for structural changes between $^{65}$Co and $^{67}$Co. A strong transition was observed in $^{65}$Co from the $1996$-keV level to the spherical $(7/2^{-})$ ground state, whereas the analogous ground state transition was not observed from any of the $\sim 2.75$-MeV levels in $^{67}$Co. The structure could not be discussed quantitatively due to the lack of reliable large-scale
shell model calculations. Nevertheless, it is clear that
low-energy proton intruder configurations are now observed from $N=38$ onwards due to the strong tensor interaction between the $\pi f^{-1}_{7/2}$ proton hole and the $\nu f_{5/2}$ and $\nu g_{9/2}$ orbitals, demonstrating how subtle the $N=40$ subshell gap is.

\begin{acknowledgments}
We gratefully thank J.~Gentens and P.~Van den Bergh for running the
LISOL separator and we acknowledge the support by the European
Commission within the Sixth Framework Programme through I3-EURONS
(contract no.~RII3-CT-2004-506065), BriX-IUAP P6/23, FWO-Vlaanderen
(Belgium), GOA/2004/03, the Foundation for Polish Science
(A.K.), the U.S. Department of Energy, Office of Nuclear Physics, under Contracts DEFG02-94ER40834 and DE-AC02-06CH11357, and the Alexander von Humboldt Foundation (W.B.W.).
\end{acknowledgments}


\begin{thebibliography}{60}
\expandafter\ifx\csname natexlab\endcsname\relax\def\natexlab#1{#1}\fi
\expandafter\ifx\csname bibnamefont\endcsname\relax
  \def\bibnamefont#1{#1}\fi
\expandafter\ifx\csname bibfnamefont\endcsname\relax
  \def\bibfnamefont#1{#1}\fi
\expandafter\ifx\csname citenamefont\endcsname\relax
  \def\citenamefont#1{#1}\fi
\expandafter\ifx\csname url\endcsname\relax
  \def\url#1{\texttt{#1}}\fi
\expandafter\ifx\csname urlprefix\endcsname\relax\def\urlprefix{URL }\fi
\providecommand{\bibinfo}[2]{#2}
\providecommand{\eprint}[2][]{\url{#2}}

\bibitem[{\citenamefont{Bernas et~al.}(1982)}]{Ber_PL_82}
\bibinfo{author}{\bibfnamefont{M.}~\bibnamefont{Bernas}} \bibnamefont{et~al.},
  \bibinfo{journal}{Phys. Lett. B} \textbf{\bibinfo{volume}{113}},
  \bibinfo{pages}{279} (\bibinfo{year}{1982}).

\bibitem[{\citenamefont{Broda et~al.}(1995)}]{Bro_PRL_95}
\bibinfo{author}{\bibfnamefont{R.}~\bibnamefont{Broda}} \bibnamefont{et~al.},
  \bibinfo{journal}{Phys. Rev. Lett.} \textbf{\bibinfo{volume}{74}},
  \bibinfo{pages}{868} (\bibinfo{year}{1995}).

\bibitem[{\citenamefont{Pawlat et~al.}(1994)}]{Paw_NPA_94}
\bibinfo{author}{\bibfnamefont{T.}~\bibnamefont{Pawlat}} \bibnamefont{et~al.},
  \bibinfo{journal}{Nucl. Phys. A} \textbf{\bibinfo{volume}{574}},
  \bibinfo{pages}{623} (\bibinfo{year}{1994}).

\bibitem[{\citenamefont{Grzywacz et~al.}(1998)}]{Grz_PRL_98}
\bibinfo{author}{\bibfnamefont{R.}~\bibnamefont{Grzywacz}}
  \bibnamefont{et~al.}, \bibinfo{journal}{Phys. Rev. Lett.}
  \textbf{\bibinfo{volume}{81}}, \bibinfo{pages}{766} (\bibinfo{year}{1998}).

\bibitem[{\citenamefont{Franchoo et~al.}(1998)}]{Fra_PRL_98}
\bibinfo{author}{\bibfnamefont{S.}~\bibnamefont{Franchoo}}
  \bibnamefont{et~al.}, \bibinfo{journal}{Phys. Rev. Lett.}
  \textbf{\bibinfo{volume}{81}}, \bibinfo{pages}{3100} (\bibinfo{year}{1998}).

\bibitem[{\citenamefont{Mueller et~al.}(1999)}]{Mue_PRL_99}
\bibinfo{author}{\bibfnamefont{W.~F.} \bibnamefont{Mueller}}
  \bibnamefont{et~al.}, \bibinfo{journal}{Phys. Rev. Lett.}
  \textbf{\bibinfo{volume}{83}}, \bibinfo{pages}{3613} (\bibinfo{year}{1999}).

\bibitem[{\citenamefont{Weissman et~al.}(1999{\natexlab{a}})}]{Wei_PRC_99}
\bibinfo{author}{\bibfnamefont{L.}~\bibnamefont{Weissman}}
  \bibnamefont{et~al.}, \bibinfo{journal}{Phys. Rev. C}
  \textbf{\bibinfo{volume}{59}}, \bibinfo{pages}{2004}
  (\bibinfo{year}{1999}{\natexlab{a}}).

\bibitem[{\citenamefont{Mueller et~al.}(2000)}]{Mue_PRC_00}
\bibinfo{author}{\bibfnamefont{W.~F.} \bibnamefont{Mueller}}
  \bibnamefont{et~al.}, \bibinfo{journal}{Phys. Rev. C}
  \textbf{\bibinfo{volume}{61}}, \bibinfo{pages}{054308}
  (\bibinfo{year}{2000}).

\bibitem[{\citenamefont{{J. Van Roosbroeck} et~al.}(2004)}]{Van_PRC_04}
\bibinfo{author}{\bibnamefont{{J. Van Roosbroeck}}} \bibnamefont{et~al.},
  \bibinfo{journal}{Phys. Rev. C} \textbf{\bibinfo{volume}{69}},
  \bibinfo{pages}{034313} (\bibinfo{year}{2004}).

\bibitem[{\citenamefont{Zeidman and {J.A. Nolen, Jr}}(1978)}]{Zei_PRC_78}
\bibinfo{author}{\bibfnamefont{B.}~\bibnamefont{Zeidman}} \bibnamefont{and}
  \bibinfo{author}{\bibnamefont{{J.A. Nolen, Jr}}}, \bibinfo{journal}{Phys.
  Rev. C} \textbf{\bibinfo{volume}{18}}, \bibinfo{pages}{2122}
  (\bibinfo{year}{1978}).

\bibitem[{\citenamefont{Georgiev et~al.}(2002)}]{Geo_JPG_02}
\bibinfo{author}{\bibfnamefont{G.}~\bibnamefont{Georgiev}}
  \bibnamefont{et~al.}, \bibinfo{journal}{J. Phys. G}
  \textbf{\bibinfo{volume}{28}}, \bibinfo{pages}{2993} (\bibinfo{year}{2002}).

\bibitem[{\citenamefont{Sorlin et~al.}(2002)}]{Sor_PRL_02}
\bibinfo{author}{\bibfnamefont{O.}~\bibnamefont{Sorlin}} \bibnamefont{et~al.},
  \bibinfo{journal}{Phys. Rev. Lett.} \textbf{\bibinfo{volume}{88}},
  \bibinfo{pages}{092501} (\bibinfo{year}{2002}).

\bibitem[{\citenamefont{Bree et~al.}(2008)}]{Bre_PRC_08}
\bibinfo{author}{\bibfnamefont{N.}~\bibnamefont{Bree}} \bibnamefont{et~al.},
  \bibinfo{journal}{Phys. Rev. C} \textbf{\bibinfo{volume}{78}},
  \bibinfo{pages}{047301} (\bibinfo{year}{2008}).

\bibitem[{\citenamefont{Stefanescu et~al.}(2007)}]{Ste_PRL_07}
\bibinfo{author}{\bibfnamefont{I.}~\bibnamefont{Stefanescu}}
  \bibnamefont{et~al.}, \bibinfo{journal}{Phys. Rev. Lett.}
  \textbf{\bibinfo{volume}{98}}, \bibinfo{pages}{122701}
  (\bibinfo{year}{2007}).

\bibitem[{\citenamefont{Stefanescu et~al.}(2008)}]{Ste_PRL_08}
\bibinfo{author}{\bibfnamefont{I.}~\bibnamefont{Stefanescu}}
  \bibnamefont{et~al.}, \bibinfo{journal}{Phys. Rev. Lett.}
  \textbf{\bibinfo{volume}{100}}, \bibinfo{pages}{112502}
  (\bibinfo{year}{2008}).

\bibitem[{\citenamefont{{P. H. Regan} et~al.}(1996)\citenamefont{{P. H. Regan},
  {J. W. Arrison}, {U. J. H{\"u}ttmeier}, and {D. P. Balamuth}}}]{Reg_PRC_96}
\bibinfo{author}{\bibnamefont{{P. H. Regan}}},
  \bibinfo{author}{\bibnamefont{{J. W. Arrison}}},
  \bibinfo{author}{\bibnamefont{{U. J. H{\"u}ttmeier}}}, \bibnamefont{and}
  \bibinfo{author}{\bibnamefont{{D. P. Balamuth}}}, \bibinfo{journal}{Phys.
  Rev. C} \textbf{\bibinfo{volume}{54}}, \bibinfo{pages}{1084}
  (\bibinfo{year}{1996}).

\bibitem[{\citenamefont{Sorlin et~al.}(2000)}]{Sor_NPA_00}
\bibinfo{author}{\bibfnamefont{O.}~\bibnamefont{Sorlin}} \bibnamefont{et~al.},
  \bibinfo{journal}{Nucl. Phys. A} \textbf{\bibinfo{volume}{669}},
  \bibinfo{pages}{351} (\bibinfo{year}{2000}).

\bibitem[{\citenamefont{Sawicka et~al.}(2004)}]{Saw_EPJ_04}
\bibinfo{author}{\bibfnamefont{M.}~\bibnamefont{Sawicka}} \bibnamefont{et~al.},
  \bibinfo{journal}{Eur. Phys. J. A} \textbf{\bibinfo{volume}{22}},
  \bibinfo{pages}{455} (\bibinfo{year}{2004}).

\bibitem[{\citenamefont{Rajabali et~al.}(2007)}]{Raj_SPR_07}
\bibinfo{author}{\bibfnamefont{M.~M.} \bibnamefont{Rajabali}}
  \bibnamefont{et~al.}, in \emph{\bibinfo{booktitle}{Proceedings of the Fourth
  International Conference on Fission and Properties of Neutron-Rich Nuclei}},
  edited by \bibinfo{editor}{\bibfnamefont{J.~H.} \bibnamefont{Hamilton}},
  \bibinfo{editor}{\bibfnamefont{A.~V.} \bibnamefont{Ramayya}},
  \bibnamefont{and} \bibinfo{editor}{\bibfnamefont{H.~K.} \bibnamefont{Carter}}
  (\bibinfo{address}{Sanibel Island, USA}, \bibinfo{year}{2007}), p.
  \bibinfo{pages}{679}.

\bibitem[{\citenamefont{Gaudefroy}(2005)}]{Gau_The_05}
\bibinfo{author}{\bibfnamefont{L.}~\bibnamefont{Gaudefroy}}, Ph.D. thesis,
  \bibinfo{school}{Universit{\'e} de Paris XI Orsay} (\bibinfo{year}{2005}).

\bibitem[{\citenamefont{Hoteling et~al.}(2006)}]{Hot_PRC_06}
\bibinfo{author}{\bibfnamefont{N.}~\bibnamefont{Hoteling}}
  \bibnamefont{et~al.}, \bibinfo{journal}{Phys. Rev. C}
  \textbf{\bibinfo{volume}{74}}, \bibinfo{pages}{064313}
  (\bibinfo{year}{2006}).

\bibitem[{\citenamefont{Bosch et~al.}(1988)}]{Bos_NPA_88}
\bibinfo{author}{\bibfnamefont{U.}~\bibnamefont{Bosch}} \bibnamefont{et~al.},
  \bibinfo{journal}{Nucl. Phys. A} \textbf{\bibinfo{volume}{477}},
  \bibinfo{pages}{89} (\bibinfo{year}{1988}).

\bibitem[{\citenamefont{Pauwels et~al.}(2008{\natexlab{a}})}]{Pau_PRC_08}
\bibinfo{author}{\bibfnamefont{D.}~\bibnamefont{Pauwels}} \bibnamefont{et~al.},
  \bibinfo{journal}{Phys. Rev. C} \textbf{\bibinfo{volume}{78}},
  \bibinfo{pages}{041307(R)} (\bibinfo{year}{2008}{\natexlab{a}}).

\bibitem[{\citenamefont{Otsuka et~al.}(2005)\citenamefont{Otsuka, Suzuki,
  Fujimoto, Grawe, and Akaishi}}]{Ots_PRL_05}
\bibinfo{author}{\bibfnamefont{T.}~\bibnamefont{Otsuka}},
  \bibinfo{author}{\bibfnamefont{T.}~\bibnamefont{Suzuki}},
  \bibinfo{author}{\bibfnamefont{R.}~\bibnamefont{Fujimoto}},
  \bibinfo{author}{\bibfnamefont{H.}~\bibnamefont{Grawe}}, \bibnamefont{and}
  \bibinfo{author}{\bibfnamefont{Y.}~\bibnamefont{Akaishi}},
  \bibinfo{journal}{Phys. Rev. Lett.} \textbf{\bibinfo{volume}{95}},
  \bibinfo{pages}{232502} (\bibinfo{year}{2005}).

\bibitem[{\citenamefont{Caurier et~al.}(2002)\citenamefont{Caurier, Nowacki,
  and Poves}}]{Cau_EPJ_02}
\bibinfo{author}{\bibfnamefont{E.}~\bibnamefont{Caurier}},
  \bibinfo{author}{\bibfnamefont{F.}~\bibnamefont{Nowacki}}, \bibnamefont{and}
  \bibinfo{author}{\bibfnamefont{A.}~\bibnamefont{Poves}},
  \bibinfo{journal}{Eur. Phys. J. A} \textbf{\bibinfo{volume}{15}},
  \bibinfo{pages}{145} (\bibinfo{year}{2002}).

\bibitem[{\citenamefont{Sorlin et~al.}(2003)}]{Sor_EPJ_03}
\bibinfo{author}{\bibfnamefont{O.}~\bibnamefont{Sorlin}} \bibnamefont{et~al.},
  \bibinfo{journal}{Eur. Phys. J. A} \textbf{\bibinfo{volume}{16}},
  \bibinfo{pages}{55} (\bibinfo{year}{2003}).

\bibitem[{\citenamefont{{N. A. Smirnova} et~al.}(2004)\citenamefont{{N. A.
  Smirnova}, {A. De Maesschalck}, {A. Van Dyck}, and Heyde}}]{Smi_PRC_04}
\bibinfo{author}{\bibnamefont{{N. A. Smirnova}}},
  \bibinfo{author}{\bibnamefont{{A. De Maesschalck}}},
  \bibinfo{author}{\bibnamefont{{A. Van Dyck}}}, \bibnamefont{and}
  \bibinfo{author}{\bibfnamefont{K.}~\bibnamefont{Heyde}},
  \bibinfo{journal}{Phys. Rev. C} \textbf{\bibinfo{volume}{69}},
  \bibinfo{pages}{044306} (\bibinfo{year}{2004}).

\bibitem[{\citenamefont{Hjorth-Jensen et~al.}(1995)\citenamefont{Hjorth-Jensen,
  {T.T.S. Kuo}, and Osnes}}]{Hjo_PR_95}
\bibinfo{author}{\bibfnamefont{M.}~\bibnamefont{Hjorth-Jensen}},
  \bibinfo{author}{\bibnamefont{{T.T.S. Kuo}}}, \bibnamefont{and}
  \bibinfo{author}{\bibfnamefont{E.}~\bibnamefont{Osnes}},
  \bibinfo{journal}{Phys. Rep.} \textbf{\bibinfo{volume}{261}},
  \bibinfo{pages}{125} (\bibinfo{year}{1995}).

\bibitem[{\citenamefont{Nowacki}(1996)}]{Now_The_96}
\bibinfo{author}{\bibfnamefont{F.}~\bibnamefont{Nowacki}}, Ph.D. thesis,
  \bibinfo{school}{IReS, Strasbourg} (\bibinfo{year}{1996}).

\bibitem[{\citenamefont{Hannawald et~al.}(1999)}]{Han_PRL_99}
\bibinfo{author}{\bibfnamefont{M.}~\bibnamefont{Hannawald}}
  \bibnamefont{et~al.}, \bibinfo{journal}{Phys. Rev. Lett.}
  \textbf{\bibinfo{volume}{82}}, \bibinfo{pages}{1391} (\bibinfo{year}{1999}).

\bibitem[{\citenamefont{Adrich et~al.}(2008)}]{Adr_PRC_08}
\bibinfo{author}{\bibfnamefont{P.}~\bibnamefont{Adrich}} \bibnamefont{et~al.},
  \bibinfo{journal}{Phys. Rev. C} \textbf{\bibinfo{volume}{77}},
  \bibinfo{pages}{054306} (\bibinfo{year}{2008}).

\bibitem[{\citenamefont{Aoi et~al.}(2008)}]{Aoi_NPA_08}
\bibinfo{author}{\bibfnamefont{N.}~\bibnamefont{Aoi}} \bibnamefont{et~al.},
  \bibinfo{journal}{Nucl. Phys. A} \textbf{\bibinfo{volume}{805}},
  \bibinfo{pages}{400c} (\bibinfo{year}{2008}).

\bibitem[{\citenamefont{Rahaman et~al.}(2007)}]{Rah_EPJ_07}
\bibinfo{author}{\bibfnamefont{S.}~\bibnamefont{Rahaman}} \bibnamefont{et~al.},
  \bibinfo{journal}{Eur. Phys. J. A} \textbf{\bibinfo{volume}{34}},
  \bibinfo{pages}{5} (\bibinfo{year}{2007}).

\bibitem[{\citenamefont{Audi et~al.}(2003)\citenamefont{Audi, Wapstra, and
  Thibault}}]{Aud_NPA_03}
\bibinfo{author}{\bibfnamefont{G.}~\bibnamefont{Audi}},
  \bibinfo{author}{\bibfnamefont{A.~H.} \bibnamefont{Wapstra}},
  \bibnamefont{and} \bibinfo{author}{\bibfnamefont{C.}~\bibnamefont{Thibault}},
  \bibinfo{journal}{Nucl. Phys. A} \textbf{\bibinfo{volume}{729}},
  \bibinfo{pages}{337} (\bibinfo{year}{2003}).

\bibitem[{\citenamefont{Otsuka et~al.}(2006)\citenamefont{Otsuka, Matsuo, and
  Abe}}]{Ots_PRL_06}
\bibinfo{author}{\bibfnamefont{T.}~\bibnamefont{Otsuka}},
  \bibinfo{author}{\bibfnamefont{T.}~\bibnamefont{Matsuo}}, \bibnamefont{and}
  \bibinfo{author}{\bibfnamefont{D.}~\bibnamefont{Abe}},
  \bibinfo{journal}{Phys. Rev. Lett.} \textbf{\bibinfo{volume}{97}},
  \bibinfo{pages}{162501} (\bibinfo{year}{2006}).

\bibitem[{\citenamefont{Hakala et~al.}(2008)}]{Hak_PRL_08}
\bibinfo{author}{\bibfnamefont{J.}~\bibnamefont{Hakala}} \bibnamefont{et~al.},
  \bibinfo{journal}{Phys. Rev. Lett.} \textbf{\bibinfo{volume}{101}},
  \bibinfo{pages}{052502} (\bibinfo{year}{2008}).

\bibitem[{\citenamefont{Oros-Peusquens and Mantica}(2000)}]{Oro_NPA_00}
\bibinfo{author}{\bibfnamefont{A.~M.} \bibnamefont{Oros-Peusquens}}
  \bibnamefont{and} \bibinfo{author}{\bibfnamefont{P.~F.}
  \bibnamefont{Mantica}}, \bibinfo{journal}{Nucl. Phys. A}
  \textbf{\bibinfo{volume}{669}}, \bibinfo{pages}{81} (\bibinfo{year}{2000}).

\bibitem[{\citenamefont{Kudryavtsev et~al.}(2003)}]{Kud_NIM_03}
\bibinfo{author}{\bibfnamefont{Y.}~\bibnamefont{Kudryavtsev}}
  \bibnamefont{et~al.}, \bibinfo{journal}{Nucl. Instr. Meth. Phys. Res. B}
  \textbf{\bibinfo{volume}{204}}, \bibinfo{pages}{336} (\bibinfo{year}{2003}).

\bibitem[{\citenamefont{Facina et~al.}(2004)}]{Fac_NIM_04}
\bibinfo{author}{\bibfnamefont{M.}~\bibnamefont{Facina}} \bibnamefont{et~al.},
  \bibinfo{journal}{Nucl. Instr. Meth. Phys. Res. B}
  \textbf{\bibinfo{volume}{226}}, \bibinfo{pages}{401} (\bibinfo{year}{2004}).

\bibitem[{\citenamefont{{P.~Van~den~Bergh} et~al.}(1997)}]{Van_NIM_97}
\bibinfo{author}{\bibnamefont{{P.~Van~den~Bergh}}} \bibnamefont{et~al.},
  \bibinfo{journal}{Nucl. Instr. Meth. Phys. Res. B}
  \textbf{\bibinfo{volume}{126}}, \bibinfo{pages}{194} (\bibinfo{year}{1997}).

\bibitem[{\citenamefont{Eberth et~al.}(2001)}]{Ebe_NPA_01}
\bibinfo{author}{\bibfnamefont{J.}~\bibnamefont{Eberth}} \bibnamefont{et~al.},
  \bibinfo{journal}{Prog. in Part. and Nucl. Phys.}
  \textbf{\bibinfo{volume}{46}}, \bibinfo{pages}{389} (\bibinfo{year}{2001}).

\bibitem[{\citenamefont{Pauwels et~al.}(2008{\natexlab{b}})}]{Pau_NIM_08}
\bibinfo{author}{\bibfnamefont{D.}~\bibnamefont{Pauwels}} \bibnamefont{et~al.},
  \bibinfo{journal}{Nucl. Instr. Meth. Phys. Res. B}
  \textbf{\bibinfo{volume}{266}}, \bibinfo{pages}{4600}
  (\bibinfo{year}{2008}{\natexlab{b}}).

\bibitem[{\citenamefont{Weissman et~al.}(1999{\natexlab{b}})}]{Wei_NIM_99}
\bibinfo{author}{\bibfnamefont{L.}~\bibnamefont{Weissman}}
  \bibnamefont{et~al.}, \bibinfo{journal}{Nucl. Instr. Meth. Phys. Res. A}
  \textbf{\bibinfo{volume}{423}}, \bibinfo{pages}{328}
  (\bibinfo{year}{1999}{\natexlab{b}}).

\bibitem[{\citenamefont{Hoteling et~al.}(2008)}]{Hot_PRC_08}
\bibinfo{author}{\bibfnamefont{N.}~\bibnamefont{Hoteling}}
  \bibnamefont{et~al.}, \bibinfo{journal}{Phys. Rev. C}
  \textbf{\bibinfo{volume}{77}}, \bibinfo{pages}{044314}
  (\bibinfo{year}{2008}).

\bibitem[{\citenamefont{Lee}(1990)}]{Lee_NPA_90}
\bibinfo{author}{\bibfnamefont{I.~Y.} \bibnamefont{Lee}},
  \bibinfo{journal}{Nucl. Phys. A} \textbf{\bibinfo{volume}{520}},
  \bibinfo{pages}{c641} (\bibinfo{year}{1990}).

\bibitem[{NND()}]{NNDC}
\bibinfo{note}{URL: \url{http://www.nndc.bnl.gov/ensdf/}}.

\bibitem[{\citenamefont{Runte et~al.}(1985)}]{Run_NPA_85}
\bibinfo{author}{\bibfnamefont{E.}~\bibnamefont{Runte}} \bibnamefont{et~al.},
  \bibinfo{journal}{Nucl. Phys. A} \textbf{\bibinfo{volume}{441}},
  \bibinfo{pages}{237} (\bibinfo{year}{1985}).

\bibitem[{\citenamefont{Flynn et~al.}(1979)\citenamefont{Flynn, Brown, Correll,
  Hanson, and Hardekopf}}]{Fly_PRL_78}
\bibinfo{author}{\bibfnamefont{E.~R.} \bibnamefont{Flynn}},
  \bibinfo{author}{\bibfnamefont{R.~E.} \bibnamefont{Brown}},
  \bibinfo{author}{\bibfnamefont{F.~D.} \bibnamefont{Correll}},
  \bibinfo{author}{\bibfnamefont{D.~L.} \bibnamefont{Hanson}},
  \bibnamefont{and} \bibinfo{author}{\bibfnamefont{R.~A.}
  \bibnamefont{Hardekopf}}, \bibinfo{journal}{Phys. Rev. Lett.}
  \textbf{\bibinfo{volume}{42}}, \bibinfo{pages}{626} (\bibinfo{year}{1979}).

\bibitem[{\citenamefont{Cochavi and Kane}(1972)}]{Coc_PRC_72}
\bibinfo{author}{\bibfnamefont{S.}~\bibnamefont{Cochavi}} \bibnamefont{and}
  \bibinfo{author}{\bibfnamefont{W.~R.} \bibnamefont{Kane}},
  \bibinfo{journal}{Phys. Rev. C} \textbf{\bibinfo{volume}{6}},
  \bibinfo{pages}{1650} (\bibinfo{year}{1972}).

\bibitem[{\citenamefont{Block et~al.}(2008{\natexlab{a}})}]{Blo_PRL_08}
\bibinfo{author}{\bibfnamefont{M.}~\bibnamefont{Block}} \bibnamefont{et~al.},
  \bibinfo{journal}{Phys. Rev. Lett.} \textbf{\bibinfo{volume}{100}},
  \bibinfo{pages}{132501} (\bibinfo{year}{2008}{\natexlab{a}}).

\bibitem[{\citenamefont{Block et~al.}(2008{\natexlab{b}})}]{Blo_PRL_08_err}
\bibinfo{author}{\bibfnamefont{M.}~\bibnamefont{Block}} \bibnamefont{et~al.},
  \bibinfo{journal}{Phys. Rev. Lett.} \textbf{\bibinfo{volume}{101}},
  \bibinfo{pages}{059901(E)} (\bibinfo{year}{2008}{\natexlab{b}}),
  \bibinfo{note}{erratum of Ref.~\cite{Blo_PRL_08}}.

\bibitem[{\citenamefont{Daugas et~al.}(2006)}]{Dau_IAP_06}
\bibinfo{author}{\bibfnamefont{J.~M.} \bibnamefont{Daugas}}
  \bibnamefont{et~al.}, \bibinfo{journal}{AIP Conf. Proc.}
  \textbf{\bibinfo{volume}{831}}, \bibinfo{pages}{427} (\bibinfo{year}{2006}).

\bibitem[{\citenamefont{Czajkowski et~al.}(1994)}]{Cza_ZPA_94}
\bibinfo{author}{\bibfnamefont{S.}~\bibnamefont{Czajkowski}}
  \bibnamefont{et~al.}, \bibinfo{journal}{Z. Phys. A}
  \textbf{\bibinfo{volume}{348}}, \bibinfo{pages}{267} (\bibinfo{year}{1994}).

\bibitem[{\citenamefont{Radford}(1995)}]{Rad_NPA_95}
\bibinfo{author}{\bibfnamefont{D.}~\bibnamefont{Radford}},
  \bibinfo{journal}{Nucl. Instr. Met. Phys. Res. A}
  \textbf{\bibinfo{volume}{361}}, \bibinfo{pages}{297} (\bibinfo{year}{1995}).

\bibitem[{\citenamefont{Bakoyeorgos et~al.}(1982)\citenamefont{Bakoyeorgos,
  Paradellis, and {P. A. Assimakopoulos}}}]{Bak_PRC_82}
\bibinfo{author}{\bibfnamefont{P.}~\bibnamefont{Bakoyeorgos}},
  \bibinfo{author}{\bibfnamefont{T.}~\bibnamefont{Paradellis}},
  \bibnamefont{and} \bibinfo{author}{\bibnamefont{{P. A. Assimakopoulos}}},
  \bibinfo{journal}{Phys. Rev. C} \textbf{\bibinfo{volume}{25}},
  \bibinfo{pages}{2947} (\bibinfo{year}{1982}).

\bibitem[{\citenamefont{Warburton et~al.}(1977)\citenamefont{Warburton, Olness,
  Nathan, Kolata, and McGrory}}]{War_PRC_77}
\bibinfo{author}{\bibfnamefont{E.~K.} \bibnamefont{Warburton}},
  \bibinfo{author}{\bibfnamefont{J.~W.} \bibnamefont{Olness}},
  \bibinfo{author}{\bibfnamefont{A.~M.} \bibnamefont{Nathan}},
  \bibinfo{author}{\bibfnamefont{J.~J.} \bibnamefont{Kolata}},
  \bibnamefont{and} \bibinfo{author}{\bibfnamefont{J.~B.}
  \bibnamefont{McGrory}}, \bibinfo{journal}{Phys. Rev. C}
  \textbf{\bibinfo{volume}{16}}, \bibinfo{pages}{1027} (\bibinfo{year}{1977}).

\bibitem[{\citenamefont{Sawicka et~al.}(2003)}]{Saw_EPJ_03}
\bibinfo{author}{\bibfnamefont{M.}~\bibnamefont{Sawicka}} \bibnamefont{et~al.},
  \bibinfo{journal}{Eur. Phys. J. A} \textbf{\bibinfo{volume}{16}},
  \bibinfo{pages}{51} (\bibinfo{year}{2003}).

\bibitem[{\citenamefont{Ivanov}(2007)}]{Iva_The_07}
\bibinfo{author}{\bibfnamefont{O.}~\bibnamefont{Ivanov}}, Ph.D. thesis,
  \bibinfo{school}{Katholieke Universiteit Leuven} (\bibinfo{year}{2007}).

\bibitem[{\citenamefont{{K. W. C. Stewart} et~al.}(1971)\citenamefont{{K. W. C.
  Stewart}, {B. Castel}, and {B. P. Singh}}}]{Ste_PRC_71}
\bibinfo{author}{\bibnamefont{{K. W. C. Stewart}}},
  \bibinfo{author}{\bibnamefont{{B. Castel}}}, \bibnamefont{and}
  \bibinfo{author}{\bibnamefont{{B. P. Singh}}}, \bibinfo{journal}{Phys. Rev.
  C} \textbf{\bibinfo{volume}{4}}, \bibinfo{pages}{2131}
  (\bibinfo{year}{1971}).

\bibitem[{\citenamefont{Heyde et~al.}(1983)\citenamefont{Heyde, {P. Van
  Isacker}, Waroquier, Wood, and Meyer}}]{Hey_PR_83}
\bibinfo{author}{\bibfnamefont{K.}~\bibnamefont{Heyde}},
  \bibinfo{author}{\bibnamefont{{P. Van Isacker}}},
  \bibinfo{author}{\bibfnamefont{M.}~\bibnamefont{Waroquier}},
  \bibinfo{author}{\bibfnamefont{J.~L.} \bibnamefont{Wood}}, \bibnamefont{and}
  \bibinfo{author}{\bibfnamefont{R.~A.} \bibnamefont{Meyer}},
  \bibinfo{journal}{Phys. Rep.} \textbf{\bibinfo{volume}{102}},
  \bibinfo{pages}{291} (\bibinfo{year}{1983}).

\end{thebibliography}
\end{document}